# Widefield two-photon random illumination microscopy (2P-RIM)

**Assia Benachir[1], Xiangyi Li[1], Eric M. Fantuzzi[1], Guillaume Giroussens[1], Thomas Mangeat[2], Federico Vernuccio[1], Hervé Rigneault[1,\*], Anne Sentenac[1,\*], and Sandro Heuke[1,\*]**

[1]*Aix Marseille Univ, CNRS, Centrale Med, Institut Fresnel, Marseille, France.*
[2]*Centre de Biologie Intégrative, Toulouse*
\**Corresponding authors: anne.sentenac@fresnel.fr, sandro.heuke@fresnel.fr, herve.rigneault@fresnel.fr*

**Abstract:** Biological and biomedical samples are routinely examined using focused two-photon (2P) fluorescence microscopy due to its intrinsic axial sectioning and reduced out-of-focus bleaching. However, 2P imaging often requires excitation intensities that can damage samples through ionization and radical formation. Additionally, the lateral resolution of 2P microscopy is lower compared to linear one-photon (1P) fluorescence microscopy. Widefield 2P microscopy, using cameras, holds promise for reducing photo-toxicity while maintaining high image acquisition rates. Widefield imaging trades the high power and short integration times of sequential single point scanning for the low power and extended integration times of parallel detection across millions of pixels. However, generating effective axial sectioning over arbitrarily large fields of view (FOVs) has remained a challenge. In this work, we introduce 2P Random Illumination Microscopy (2P-RIM), an easy-to-implement 2P widefield technique, that achieves low photo-damage, fast imaging, micrometric axial sectioning, and enhanced lateral resolution for arbitrarily large FOVs. By using widefield speckled illuminations in conjunction with an image standard deviation matching algorithm, 2P-RIM demonstrated multicolor imaging over FOVs greater than 200 µm, lateral resolution 220 nm, axial sectioning 2µm, and peak excitation powers about 10 times lower than those used in focused laser scanning microscopy.

## 1. Introduction

Two-photon (2P) excitation fluorescence microscopy was developed to circumvent intrinsic and important limitations of one-photon (1P) microscopy, essentially sensitivity to light scattering, low penetration depth, and out-of-focus background [1, 2]. 2P microscopy uses near-infrared excitation light that minimizes scattering in biological tissues and the two-photon excitation strongly reduces out-of-focus background signal. Altogether, 2P microscopy allows for a better imaging depth than 1P microscopy with visible light illumination, which is of primary importance for imaging thick samples like mammalian embryos, neural tissues or optical biopsies [3]. In practice, 2P microscopy involves scanning a near-infrared focused laser beam inside the sample and detecting the visible fluorescence on a bucket detector. The nonlinearity of the excitation process generates signal photons mainly from the high-intensity region, producing optically sectioned images. However, the high power density required to obtain a detectable fluorescent signal from the tiny 2P excitation volume can cause potentially damaging multi-photon absorption events [4–6]. The phototoxicity limits the scanning speed as the laser intensity cannot be increased to compensate for the diminution of the detector integration time, which can be a handicap to image large fields of view. Another weakness of 2P scanning microscopy is its low transverse resolution, which is, at best, comparable to that of a standard 1P fluorescence microscope [7–9].

With the availability of increasingly powerful lasers, widefield configurations have emerged in the past fifteen years as faster and less phototoxic alternatives to scanning approaches. To keep an optical sectioning capability, widefield two-photon microscopes are generally coupled to a temporally focused illumination [10–12] that confines the 2P excitation to a slice a few microns thick. Importantly, quasi all practical implementations of widefield 2P microscopy rely on temporal focusing, as illuminating with a plane wave over large fields of view can lead to significant energy deposition at the back focal plane of the objective, risking optical damage. TF circumvents this issue by angularly dispersing the spectral components of the excitation pulse, and is currently the only viable solution for achieving axial confinement in widefield 2P excitation. TF-based 2P microscopes have also been combined with super-resolution methods (structured illumination, photo-activation, blinking) to improve their transverse resolution beyond that of classical 1P microscopy [13–16]. However, the temporal focusing technique, not to mention its combination with a super-resolution method, is challenging experimentally. These challenges include grating blaze efficiency limitations for multicolor excitation, sensitivity to pulse width variations, phase distortions in inhomogeneous samples, and the risk of objective lens damage for large field of views due to power throughput constraints. As a result, super-resolved images have only been demonstrated on small fields of view (less than 50×50 µm$^2$) and never for multi-color imaging. In this work, we simplify the experimental implementation of widefield two-photon microscopy by using random speckled illuminations and variance processing to obtain optical sectioning and super-resolution [17]. We show that two-photon speckle excitation reduces the out-of-focus blur and allows imaging at a deeper penetration depth than its one-photon counterpart. The potential of 2P Random Illumination Microscopy (2P-RIM), is demonstrated with super-resolved, multi-color images of various biological tissues on large FOVs (200×200 µm$^2$).

## 2. Results

### 2.1. Adapting RIM to two-photon excitation

Random illumination microscopy (RIM) is a recent super-resolution microscopy technique that was first developed for 1P fluorescence excitation [17–21]. RIM requires recording multiple images of the sample under random speckled illuminations (obtained by passing a laser beam through a diffuser), see Fig. 1a. It has been shown that the standard deviation of the low-resolution speckled images yields an optical sectioning comparable to that of a confocal microscope [22]. In addition, it was mathematically demonstrated that, under certain conditions, there is a bijection between the spatial frequencies of the variance of the speckled images and those of the sample fluorescence density over a Fourier support twice as large as that of the microscope point spread function [23]. In practice, the sample is estimated numerically by minimizing the difference between the experimental and theoretical standard deviations. RIM has been shown experimentally to produce super-resolved images with twice the resolution of standard fluorescence images, similar to Structured Illumination Microscopy (SIM) [17, 20, 24]. Importantly, RIM data processing only requires the knowledge of the speckles' statistics which are well known and very robust to aberrations, scattering, or misalignments [25]. This fundamental property makes RIM much easier to use than SIM which requires knowledge of each illumination pattern [17, 20, 24].

Adapting RIM to two-photon microscopy requires studying the statistics of the 2P speckle excitation function which differs from the 1P one. In 2P microscopy, the emitted fluorescence depends on the square intensity of the illumination while in 1P microscopy it depends on the intensity [26]. The experimental images of the 1P and 2P speckle excitations in Fig. 1c illustrate their important statistical differences.

To model the theoretical standard deviation of low-resolution images obtained under two-photon random speckled illumination, we assume that the sought fluorescence density is confined in a thin slice at the focal plane, $\rho(\mathbf{r}, z) = \rho(\mathbf{r})\delta(z)$ (where $\mathbf{r}$ is the coordinate in the transverse plane). In this case, the 2P-excited fluorescence intensity recorded by the camera for a given speckled illumination reads,

$$I_{2P}(\mathbf{R}) = \int H(\mathbf{R} - \mathbf{r})\rho(\mathbf{r})S^2(\mathbf{r})d\mathbf{r} + B(\mathbf{R}) + \epsilon(\mathbf{R}) \quad (1)$$

where $\mathbf{R}$ and $\mathbf{r}$ are coordinates in the image and object space, respectively, $H$ represents the incoherent collection point spread function at the emitted (visible) wavelength (it is similar to that used in 1P fluorescence microscopy), $S^2$ stands for the square intensity of the speckled illumination, $B$ accounts for the out-of-focus fluorescence which is assumed to be independent of the speckled illuminations and $\epsilon$ models the noise (essentially Poisson noise). Then, the variance of the speckled images (obtained with an infinite number of speckled illuminations) reads,

$$\sigma^2_{\text{model}}[\rho](\mathbf{R}) = \iint H(\mathbf{R} - \mathbf{r}_1)\rho(\mathbf{r}_1)\Lambda_{2P}(\mathbf{r}_1 - \mathbf{r}_2)\rho(\mathbf{r}_2)H(\mathbf{R} - \mathbf{r}_2)d\mathbf{r}_1 d\mathbf{r}_2 + \sigma^2_\epsilon \quad (2)$$

where $\sigma^2_\epsilon$ is the noise variance and $\Lambda_{2P}(\mathbf{r}_1 - \mathbf{r}_2) = <S^2(\mathbf{r}_1)S^2(\mathbf{r}_2)> - <S^2(\mathbf{r}_1)><S^2(\mathbf{r}_2)>$ with $<>$ standing for the average over the speckled illuminations, is the auto-covariance of the square speckled light intensity. The illumination is assumed to be a fully developed speckle that fills the objective back focal plane. In this case, it is shown (see Supplementary Note 1) that

$$<S>(\mathbf{r}) = H_{\text{ex}}(0)$$
$$<S(\mathbf{r})S(\mathbf{r}')> - <S(\mathbf{r})><S(\mathbf{r}')> = H^2_{\text{ex}}(\mathbf{r} - \mathbf{r}')$$
$$<S^2>(\mathbf{r}) = 2H^2_{\text{ex}}(0) \quad (3)$$
$$\Lambda_{2P}(\mathbf{r} - \mathbf{r}') = 4H^2_{\text{ex}}(0)H^2_{\text{ex}}(\mathbf{r} - \mathbf{r}') + 16H^4_{\text{ex}}(\mathbf{r} - \mathbf{r}')$$

$$(4)$$

where $H_{\text{ex}}$ is the incoherent point spread function of the microscope at the excitation (near-infrared) wavelength.

Once the detection and excitation point spread functions $H$ and $H_{\text{ex}}$ are known, one can simulate the image variance Eq. (2) for any fluorescence density $\rho$. The reconstruction algorithm, algoRIM consists in estimating $\rho$ that minimizes the cost functional $\mathcal{F}$,

$$\mathcal{F}[\rho] = \int \|\hat{\sigma}(\mathbf{R}) - \sigma_{\text{model}}[\rho](\mathbf{R})\|^2 + \mu|\rho|^2(\mathbf{R})d\mathbf{R}, \quad (5)$$

where $\hat{\sigma}$ is the experimental standard deviation and µ is a Tikhonov regularisation parameter (tuned manually). The key features of algoRIM lay in the fast calculation of the quadruple integral of $\sigma_{\text{model}}[\rho]$, Eq. (2), the removal of the noise beyond the optical transfer function, and a crude estimation of the noise variance. All the details of the inversion procedure are presented in the in Supplementary Note 1 as well as in [27]. The link towards the 2P-RIM user-friendly reconstruction application, together with a description of its interface is given in Fig. S1 of the Supplementary Note 4.

### 2.2. 2P-RIM brings lateral super-resolution and optical sectioning

The principles and experimental implementation of 2P-RIM are described in Fig. 1(a,b). A low repetition rate near-infrared pulsed laser was introduced in a standard widefield microscope in epi-fluorescence configuration. The speckled illuminations were formed by passing the collimated laser beam through a Spatial Light Modulator (SLM) placed at a plane conjugated to the backfocal plane of the objective. The SLM displayed random phase masks at rates between 10 to 60Hz. Depending on the sample, four hundred to two thousand speckled 2P fluorescence images were recorded on the camera.

In addition to the RIM reconstruction, we calculated the average of the speckled images (hereafter called 2P-average image). As the mean of the speckled illuminations is approximately uniform, the 2P-average image corresponds to the image obtained under homogeneous illumination. It is similar to the image of a standard 1P widefield microscope. In the following, the 2P-average image is used as a basis for comparison to investigate the improvement in lateral resolution and optical sectioning of 2P-RIM.

We first tested the performances of 2P-RIM on calibrated samples and synthetic data (see Supplementary Note 4 and 5 ). Experiments on beads of diameter 200 nm were conducted using an objective microscope with an effective Numerical Aperture (NA) of about 0.8 and with the emission and excitation wavelengths of 560 nm and 750 nm, respectively. We estimated the Full Width at Half Maximum (FWHM) of the point spread function of the 2P-average image to about 363 nm while that of RIM reconstruction was 216 nm, see Fig. S2 (a,b) of the Supplementary Note 5. A similar resolution gain of about 1.7 was found by simulating the 2P-average and the 2P-RIM images of a resolution target, see Fig. S3b of the Supplementary Note 5. The optical sectioning of RIM, due to the standard deviation processing [22], was investigated by translating a homogeneous fluorescent plane through the focal plane. It was evaluated experimentally to 2 microns in agreement with the simulations, see Figs. (S2c,S3b) of the Supplementary Note 5.

### 2.3. 2P-RIM provides multicolor super-resolved images of biological tissues on large fields of view.

To illustrate the ease of use and performance robustness of 2P-RIM, we imaged biological tissues displaying different labeling characteristics (dense or sparse, high or low fluorescence background) on large FOVs (greater than 100 microns).

In Fig. 2a, we considered immuno-stained U2O2 cells where the fluorescence of Alexa Fluor 568 phalloidin outlines the location of F-actin. As expected, the contrast and lateral resolution are significantly better in the RIM reconstruction than in the 2P-average image. In particular, two filaments 300 nm apart can be distinguished in the former whereas they are fused in the latter (see the zooms of Fig. 2a).

In Fig. 2b we imaged the autofluorescence of the chromoplasts inside a fern leaf. Plants are difficult to image using standard microscopy due to the weak penetration of visible light into pigmented tissues [3]. 2P-RIM image shows an improved resolution and reduction of out-of-focus light and noise background compared to the 2P-average image. In particular, it enables the visualization of the grana in the chloroplasts [28]. Note that the chloroplast fluorescence intensity is dimmed along contours that evoke cell membranes. These shadows indicate a localized absorption of the emitted light, likely due to a cell layer placed above the focal plane. The rapid variation of the absorption is a difficulty for 2P scanning microscopy, as the high power density of the focused beam can induce local tissue damage in absorbing regions. In this case, 2P-RIM and its inherent lower phototoxicity are particularly interesting.

To demonstrate 2P-RIM compatibility with tunable excitation laser sources, we highlight within Fig. 2 multi-color 2P-RIM imaging of the epidermis of a drosophila larva. The sample was illuminated sequentially at 750 nm and 800 nm while the fluorescence detection windows were selected to collect predominantly the fluorescence of Alexa Fluor 568 or Alexa Fluor 488. The Alexa Fluor 568 dye was linked to phalloidin binding to actin filaments while Alexa Fluor 488 was linked to a secondary anti-body outlining the distribution of primary anti-body labeled cadherines. Cadherines are glycoproteins that play a role in cell-cell adhesion and contribute to tissue differentiation. Whatever the fluorescent labels, the 2P-RIM images are significantly better resolved and contrasted than the 2P-average images, see Fig. 2. Their transverse resolution is sufficient to reveal the fibrous structures forming the actin cytoskeleton.

### 2.4. 2P-RIM can image deeper than 1P-RIM

In the previous sections, we have shown the capacity of 2P-RIM to provide widefield super-resolved and optically sectioned images of various biological samples. The essential requirement for RIM robust data processing to succeed is that the variation of the recorded data (due to the speckle fluctuation) be greater than the unavoidable variation due to noise [17]. Now, noise can rapidly become an issue, especially when imaging deep into thick or densely labeled samples. Indeed, in RIM, the whole sample is illuminated by light speckles. The fluorescence density excited by the speckle grains at the focal plane produces crisp bright spots. In contrast, the fluorescence excited by out-of-focus speckle grains produces large blurred spots, the superposition of which tends to form a quasi-homogeneous high background (noted $B$ in Eq. (1)). The intensity of the background is roughly proportional to the spatial average (which is equivalent to the ensemble average) of the fluorescence excitation function, either $<S>$ in 1P-RIM or $<S^2>$ in 2P-RIM. The Poisson noise of the background is thus proportional to $<S>$ in 1P-RIM or $<S^2>$ in 2P-RIM. In contrast, the variance of the bright in-focus spots due to the speckles' fluctuation, is proportional to the variance of the speckled excitation which is, according to Eq.(4), $Var[S] =<S^2> - <S>^2=<S>^2$ in 1P-RIM and $Var[S^2] =<S^4> - <S^2>^2= 10<S^2>^2$ in 2P-RIM. Thus, the signal-to-noise ratio is 10 times bigger in 2P-RIM than in 1P-RIM, and we expect 2P-RIM to be better adapted to the observation of dense or thick samples than 1P-RIM.

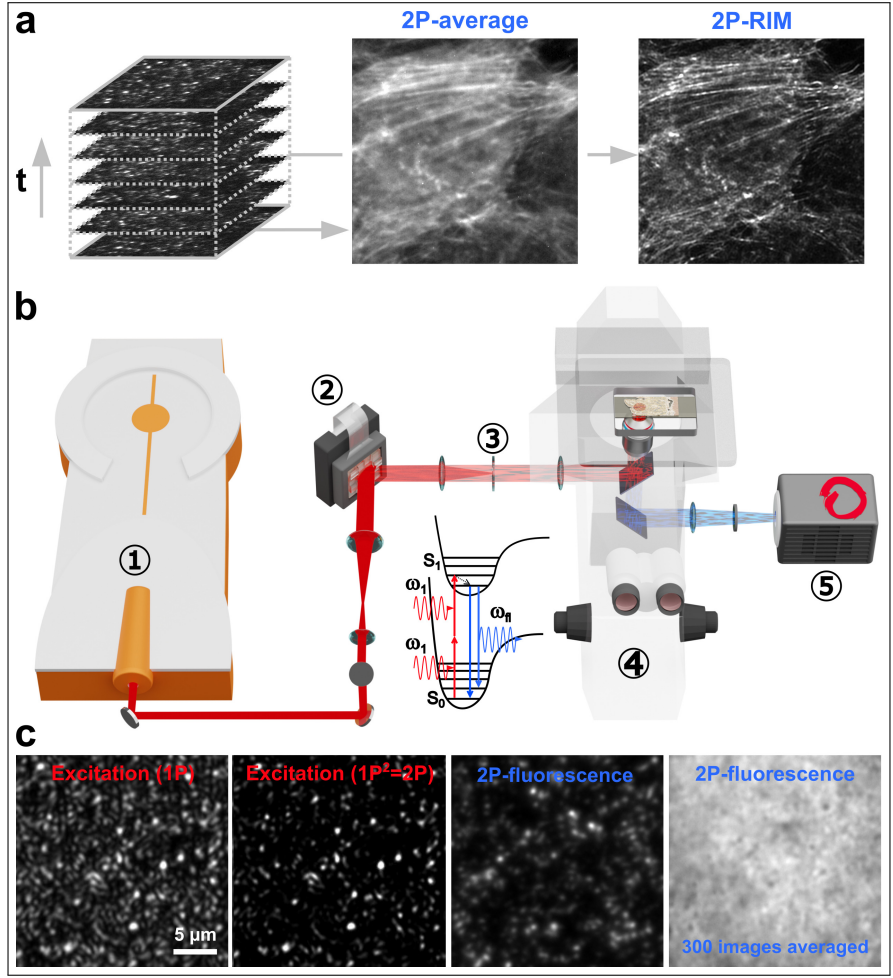

Fig. 1. Principles of 2P-RIM and Experimental setup: **a)** (left) A stack of low-resolution 2P-fluorescence images under random speckled illuminations is acquired. (center) The average of these speckled images, named the 2P-average, corresponds to the image obtained under homogeneous illumination and serves as a basis for comparison. (right) The iterative RIM reconstruction algorithm processes the standard deviation of the speckled images to yield an estimation of the sample with optical sectioning and enhanced transverse resolution. **b)** Experimental set-up. (1) A pulsed laser beam (low repetition rate fs-laser, 20W, 100kHz, 350fs, 1030nm) goes through a (2) diffuser (Spatial light modulator displaying random patterns, combined with (3) a zero-order stop) placed at a conjugated back-focal plane of the objective of a (4) standard microscope in epi-configuration. The 2P fluorescence is collected on a (5) scientific camera. **c)** (left) speckled illumination intensity as reflected by a mirror placed at the object focal plane ($\lambda_{ex}$ = 1030 nm). (center left) square of the speckled illumination intensity, corresponding to the 2P speckled excitation. (center right) Image of a homogeneous fluorescent sample under the 2P speckled excitation (emission at 530 nm). (right) The sum of 300 2P-speckled images shows the quasi-homogeneity of the 2P-speckled excitation average.

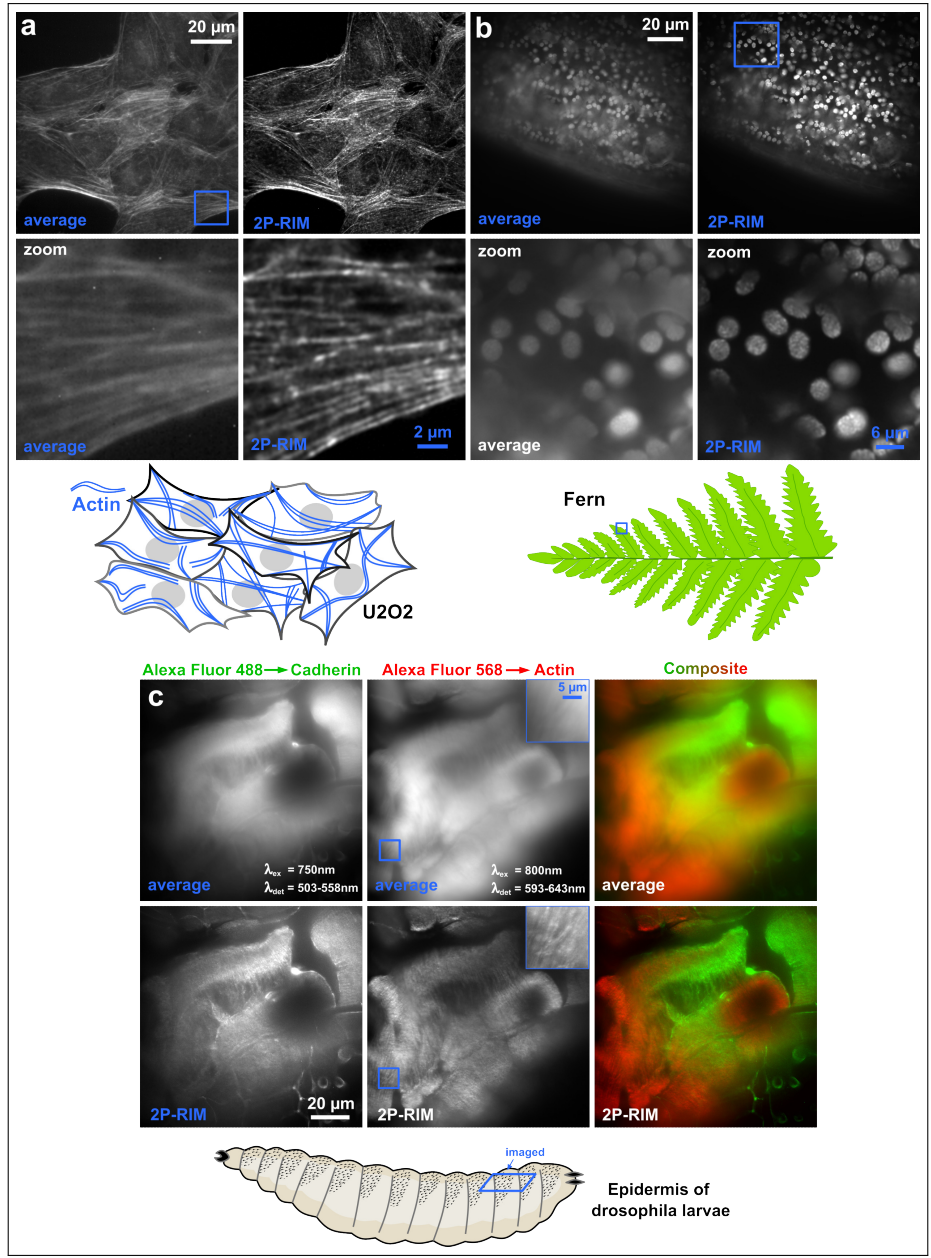

Fig. 2. Comparison of 2P-average and 2P-RIM images of fixed biological samples: **a)** Images of actin labeled (Alexa Fluor 568 phalloidin) U2O2 cells illuminated at $\lambda_{ex}$ = 1030 nm. 500 raw speckled images were recorded. In the zooms, we can see that 2P-RIM can distinguish several actin filaments in places where 2P-average shows only one large filament. **b)** Images of the autofluorescence of a fern leaf illuminated at $\lambda_{ex}$ = 1030 nm. 500 raw speckled images were recorded. Contrary to 2P-average, 2P-RIM enables the visualization of the grana inside the chromoplasts. **c)** Bicolor images of actin labeled (Alexa Fluor 568 phalloidin) and cadherin labeled (Alexa Fluor 488) epidermis of a *Drosophila* larva. Left: 2P-average (top) and 2P-RIM (bottom) images of the cadherin label. We recorded 400 speckled images with excitation wavelength $\lambda_{ex}$ = 750 nm and detection wavelength $\lambda_{det} \in [503, 558]$. Middle: Same for the actin label. We recorded 400 speckled images with $\lambda_{ex}$ = 800 nm and $\lambda_{det} \in [593, 643]$. Right: Superimposed 2P-average (top) and 2P-RIM (bottom) images of the cadherin and actin. The improvement brought by 2P-RIM in terms of contrast and resolution is significant.

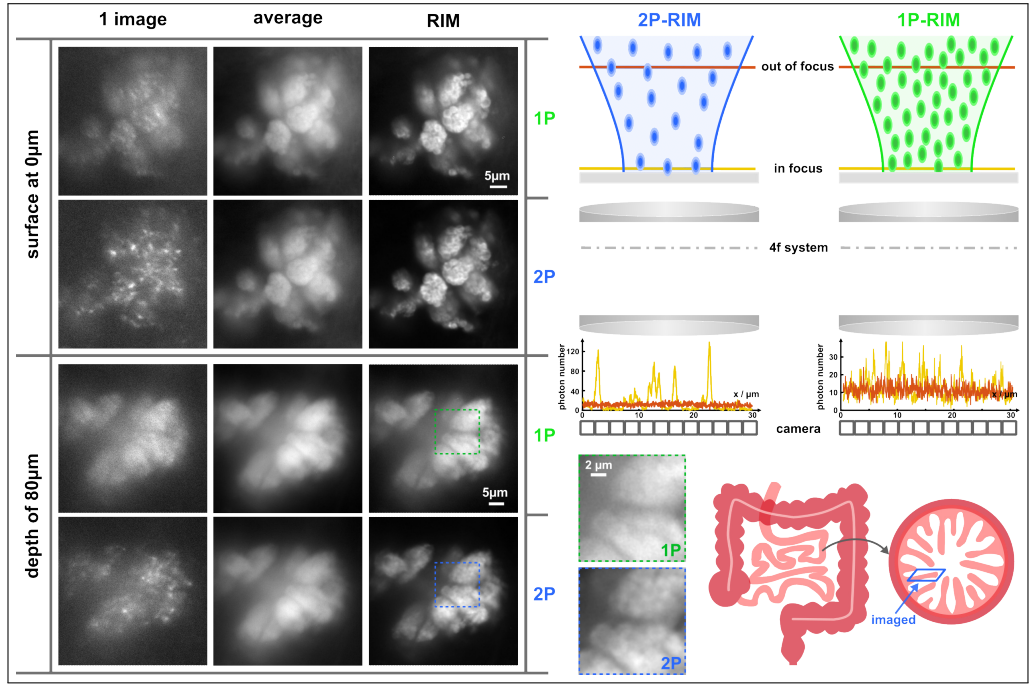

Fig. 3. Comparison of 1P-RIM and 2P-RIM at two different depths of a mouse intestinal jujunum (Left). The rhodamine-marked sample is illuminated at 532nm and 1030nm for 1P and 2P excitation respectively. 2000 speckled images are recorded with the same global photon budget for both the 1P and 2P experiments. The left column shows raw 1P and 2P speckled images. The speckle bright spots stand out better from the noisy background in the 2P excitation mode. This observation is confirmed numerically on the right-handed illustration where the fluorescent signal stemming from two fluorescent planes (one in focus and one out of focus) under 1P or 2P speckled excitation is simulated for the same photon budget and deteriorated with Poisson noise. The ratio between the in-focus signal (bright peaks) and out-of-focus signal (noisy background) is significantly higher in the 2P excitation mode than in the 1P mode. The middle column displays the 1P and 2P average images. Their similarity points out the good homogeneity of the average speckled illumination and the similar photon budget of both experiments. The right column displays the 1P and 2P RIM reconstructions. 2P-RIM is better resolved and contrasted than 1P-RIM, in agreement with its better signal-to-noise ratio.

We illustrate this analysis by imaging a rhodamine-marked mouse intestinal villi sample with 1P-and 2P-RIM (excitation wavelengths, 532 and 1030 nm respectively). In Fig. (3), we display one raw 2P and 1P speckled images, the 2P and 1P averages, and the 2P and 1P RIM reconstructions at two different depths inside the sample (close to the sample surface and 80 microns deep inside the sample). First, we observe that the in-focus bright spots stand out better from the background in the 2P raw image than in the 1P one. Second, the 1P and 2P averages are similar, indicating that the number of speckled illuminations was sufficient in both the 1P and 2P experiments to ensure a homogeneous excitation of the fluorescence. Last, 2P-RIM reconstructions are better contrasted and resolved than the 1P-RIM ones. The interest of 2P-RIM is particularly visible when imaging deep inside the sample where the observation point spread function widens due to aberration and scattering. With the in-focus bright spots being spread on a larger domain, the signal variance is diminished in both the 1P and 2P cases. It remains sufficiently high compared to the background noise in the 2P configuration but not in the 1P one. At this depth, 1P-RIM yields an image similar to 1P-average while 2P-RIM still improves the contrast and resolution of the 2P-average.

## 3. Discussion

2P-RIM is a widefield 2P-fluorescence microscopy method that involves recording multiple images of the samples under random speckled illuminations and reconstructing numerically a super-resolved image from the standard deviation of the raw low-resolution speckled images, Fig. (1). RIM reconstruction procedure requires only the knowledge of the observation point spread function and of the speckle statistics. The latter being quasi-insensitive to misalignment, aberrations, and scattering, 2P-RIM is easy to implement experimentally and numerically, see Fig. 1. The speckles are formed by passing the laser beam through a rotating diffuser or a spatial light modulator displaying random patterns. Importantly, and contrary to microscopes using temporal focusing or standard structured illumination, changing the excitation wavelength does not require any modification of the set-up, which eases significantly the acquisition of multi-color images. An additional benefit of RIM's speckled illumination is that it never focuses on a single point or single line along its optical path, reducing the risk of damage to the optical components. Hence, the peak power of the laser (and consequently the field of view) can be more easily increased in 2P-RIM than in standard widefield 2P microscopes using or not temporal focusing. As a result, 2P-RIM is the first microscope, to our knowledge, that can provide widefield multi-color super-resolved 2P-images over fields of view about 200x200 $\mu m^2$, Fig. 3.

Besides its remarkable simplicity, 2P-RIM displays critical assets in terms of resolution, optical sectioning, in-depth imaging, and large fields of view: (i) The lateral super-resolution is obtained thanks to the structured speckled illumination which down-modulates inaccessible high spatial frequencies of the sample within the optical transfer function [23]. We have shown that 2P-RIM improves the lateral resolution by about 1.7 compared to standard 1P microscopy, see Fig. 2 and Supplementary Figs. S2, S3. (ii) Optical sectioning is obtained when taking the standard deviation of the speckled images. Indeed, the out-of-focus fluorescence background varies less with the speckle fluctuations than the in-focus signal. 2P-RIM produces optical sectioning close to that of a confocal microscope [22], see Supplementary Figs. S2, S3 . (iii) Importantly, 2P-RIM proved to be more adapted than 1P-RIM for probing deep inside thick or densely labeled samples as illustrated in Fig. 3. In this example, the raw speckled images of 1P-RIM were plagued by a strong out-of-focus fluorescence background that masked the in-focus signal. As a result, the image variance was dominated by the noise variance stemming from the background and the 1P-RIM image was comparable to the 1P-average. For the same sample, the out-of-focus fluorescence background in 2P-RIM was significantly lower than the in-focus signal, and RIM reconstruction improved the 2P-average image. We have established theoretically that the ratio between the signal variance and the background noise variance is one order of magnitude higher in 2P-RIM than in 1P-RIM. Note that the background noise can be further decreased in 2P-RIM by slightly focusing the speckled illumination to reduce the axial extension of the two-photon excitation volume, (see the discussion and Fig. S3b of the supplementary information). Nevertheless, it is important to note that scattering within the sample still imposes practical limitations on 2P-RIM performance. While the excitation speckle pattern is statistically robust to aberrations and scattering, strong scattering enlarges the excited sample volume which may reduce the peak intensity. Then, higher input power is required for equivalent excitation. Additionally, scattering of the emitted fluorescence broadens the detection point spread function, leading to decreased resolution and contrast in the final RIM reconstruction. (iiii) As a widefield technique that simultaneously addresses millions of sample points, 2P-RIM has the potential for gentle and fast imaging over large areas. In this work, we used a 5-Watt laser with 200 kHz repetition rate able to trigger 2P fluorescence on fields of view up to $200 \times 200$ µm$^2$. A detailed analysis in Supplementary Note 2 shows that, for the same global fluorescence photon budget over a $200 \times 200$ µm$^2$ FOV, the acquisition time can be one hundred times faster using widefield 2P microscopy than standard 2P scanning configuration if the laser peak powers are the same. Conversely, if we keep the same acquisition times, the peak power (thus the phototoxicity) can be ten times smaller in the widefield configuration than in the scanning one. In addition, we have estimated that the heating in the widefield configuration is reduced by a factor of two compared to the scanning approach when the sample is thin enough (a few tens of microns) to allow an efficient heat dissipation toward the sample holder - see Supplementary Note 3.

However, the present implementation of 2P-RIM suffers from several limits and there is still room for improvement. Since speckled illuminations trigger the fluorescence over the whole sample volume, 2P-RIM is expected to be more phototoxic than widefield 2P microscopy using temporal focusing especially if a z-scan of the sample is performed. In this case, imaging several planes simultaneously with a remote focusing device could be a solution to reduce 2P-RIM phototoxicity and acquisition time [29, 30]. Also, 2P-RIM requires hundreds of images (400 to 2000) for the speckles' average and variance to be roughly homogeneous throughout the sample. Our experiment used a 60 Hz Spatial Light Modulator to form the speckled illuminations. The recording of the speckled images stack took from 8 to 120 seconds. For comparison, the acquisition time of a 200×200 µm$^2$ FOV in standard scanning microscopy is about 16 seconds ($4000 \times 4000$ pixels of size 50 nm with 1 µs dwell time). Binary Spatial Light Modulators at 500 Hz [17] together with an increase in the laser peak power could be used to fasten the acquisition of the images' stack. Another possibility would be to use fewer illuminations by optimizing the speckle patterns for their empirical variance to be homogeneous [31].

Taken together, 2P-RIM fills a unique niche in the landscape of advanced fluorescence microscopy by combining widefield acquisition, nonlinear sectioning, and resolution enhancement in a compact and robust setup. Its strengths become most evident in scenarios that require moderate lateral resolution (200–400 nm), low phototoxicity, and imaging depths up to 150 µm - conditions common in large, slowly evolving biological samples such as Drosophila larvae, C. elegans, or cell colonies. In addition to improved penetration depth, 2P-RIM offers the possibility of reducing heat load compared to one-photon methods: by adjusting the phase pattern on the SLM, the excitation can be spatially confined to a well-defined 3D volume. This restricts energy deposition and photobleaching to the actual imaging volume, which is particularly advantageous for scattering or thermally sensitive tissues.

We numerically confirm that the theoretical resolution enhancement of 1P-RIM and 2P-RIM is a factor of two in the ideal case. In practice, we achieve a gain of approximately 1.7 (see Supplementary Fig. S2), which is consistent with earlier 1P-RIM results. A comprehensive benchmarking of 1P-RIM against established super-resolution techniques—including STED, STORM and SIM—was published by Mangeat et al. (Cell Reports Methods, 2021, Ref. [17]) and serves as a reference for evaluating the performance of RIM-based approaches.

Compared to other super-resolution techniques, 2P-RIM offers a favorable trade-off between resolution, acquisition speed and implementation complexity. Methods such as STED or PALM/STORM offer higher spatial resolution, but typically involve demanding instrumentation, long acquisition times, or strong sensitivity to sample-induced aberrations. SIM, while conceptually more closely related to RIM, performs poorly in scattering tissues and is inherently incompatible with full-field two-photon excitation due to the intense peak power required at the back aperture of the objective. To date, no practical implementation of widefield 2P-SIM has been reported.

In contrast, 2P-RIM is tolerant to optical aberrations, does not require pattern calibration, and scales naturally to multicolour imaging and large fields of view. With future advances—such as low-repetition-rate lasers, faster speckle modulation and high-speed cameras—2P-RIM may enable multi-Hertz imaging over megapixel-scale areas. This positions 2P-RIM as a competitive alternative to scanning-based two-photon microscopy, particularly in applications requiring gentle, high-resolution imaging across large biological regions.

**Materials and methods**

*3.1. Experimental setup*

Figure 1 presents schematically our 2P-RIM experimental setup. The excitation source is a high peak power amplified Yb laser (Tangerine, Amplitude, 20W) that emits 350 fs pulses at 1030 nm with a repetition rate of 100kHz. For the demonstration of 2P-RIM's compatibility with tunable light sources in Fig. 2, we used the output of an optical parametric amplifier (OPA) which can be tuned between 700-900nm and features a pulse width of 0.8 ps at 200kHz repetition rate [32]. The laser beam (Yb-laser or OPA) is expanded to match the area of our spatial light modulator (SLM, X15213 series, Hamamatsu). Distinct speckle patterns are generated by projecting random phase masks on the SLM with 0-$2\pi$ phase shifts and a maximum refresh rate of 60Hz. As an alternative, rotating diffusers can be used, but continuous motion leads to intra-frame speckle blur, while stepwise operation is limited by mechanical inertia. These constraints reduce resolution or speed, which is why we opted for an SLM-based implementation. The SLM is imaged by a telescope to the back focal plane of the objective lens (CFI APO LWD NIR, 40×, Nikon, water immersion objective, 1.15 NA). The zeroth-order reflection from the SLM is blocked at the Fourier plane of the telescope by a custom silver mirror of 1mm diameter that was coated onto a large transparent glass substrate. The objective lens focuses the speckle pattern at the object plane with a diameter of 250µm (FWHM, 2P fluorescence intensity) that delimits the FOV. The sample is moved in the lateral direction by a motorized 2-axis microscope stage (SCANplus 100 x 100, Märzhäuser) while the axial imaging plane is adjusted using a 1-axis motorized stage (L-836.501212, PI) that lifts the objective lens. At the sample, the average excitation power does not exceed 250mW corresponding to peak power levels of <150W/µm$^2$ compared to power levels that are typically 1-2 orders of magnitude higher in focused laser scanning 2P microscopy [33]. The 2P-excited fluorescence light is collected in the backward direction by the same objective lens and imaged via a second lens (Thorlabs, ACT508-300-A-ML) and, hence, 60x magnification onto a scientific CMOS camera (ORCA-fusion, C14440-20UP, Hamamatsu). The remaining excitation light after the dichroic mirror (Thorlabs, DMSP805R) is filtered by a short-pass filter (Thorlabs, FESH0700). Considering the magnification, a maximum field of view (FOV) of 250µm × 250µm is achieved and sampled over 2304 × 2304 pixels. Stacks of speckle illuminated 2P fluorescence images are acquired using Matlab (Version R2018b, MathWorks Inc.) as control software for triggering the SLM pattern generation as well as for collecting 2P images. 400 to 2000 speckled images were used for the RIM reconstructions, corresponding to a total acquisition time between 8s and 120s per high-resolution 2P-RIM image.

### 3.2. Numerical simulations and RIM reconstruction

The numerical simulations and the principles of RIM inversion procedure are detailed in the Supplementary Note 1.

### 3.3. Samples' preparation

Fig. 2: (a) Actin stained U2O2 cells: 2P-RIM imaging was performed in U2OS osteosarcoma cells. U2OS cells were maintained in McCOY's medium (Gibco$^{TM}$, 16600082) supplemented with 10% fetal bovine serum (Dominique Dutscher, S181H), 100 U/mL penicillin and 100 µg/mL streptomycin (Sigma-Aldrich, P4333) in the presence of 5% $CO_2$ humidified atmosphere at 37ºC. Cells were seeded onto sterilized glass coverslips and allowed to attach and spread for 16 h, then fixed for 15 min using 4% paraformaldehyde (Electron Microscopy Sciences, 15714) in cytoskeleton buffer (10mM MES pH6.1, 150mM NaCl, 5mM EGTA, 5mM glucose, 5mM MgCl2). The cells were permeabilized and blocked in phosphate-buffered saline (PBS) containing 0.1% saponin and 1% bovine serum albumin (BSA) for 2 h at room temperature. Labelling of F-actin was achieved with Alexa Fluor 568 phalloidin (Thermo Fisher Scientific, A12380) after 2 h of incubation at 165 nM in 0.1% saponin/1% BSA/PBS. For 2P-RIM imaging, coverslips with stained cells were mounted with fluormount aqueous mounting medium (Sigma-Aldrich, F4680) and kept at 4°C until acquisition.
(b) Fern: We prepared the plant samples by placing a small leaf between two thin coverslips, in a drop of milliQ water. The plant is a Boston fern (Nephrolepis exaltata). No further preparation was required, the leaf was neither cut nor stained.
(c) Drosophila larva epidermis: The dissected larva epidermis was washed in PBS, fixed in 4% PFA for 30 minutes before rinsing with 1% Triton. Labelling of F-actin was achieved with Alexa Fluor 568 phalloidin (Thermo Fisher Scientific, A12379) after 2 h of incubation at 165 nM in 0.1% saponin/1% BSA/PBS. E-cadherin was labelled using antibodies (primary: Rat anti-Ecad, DSHB DCAD2; secondary: anti-rat Alexa Fluor 488).
Fig. 3: Mouse intestinal villi: Jejunum from one sacrificed mouse was fixed in 4% paraformaldehyde (Electron Microscopy Sciences), washed in PBS, sectioned at 200-300 $\mu$M thickness, permeabilized with 1% Triton X-100 for 1 h. Saccharides in the mucus were stained with wheat germ agglutinin (WGA), rhodamine, washes in PBS and mounting in Aqua-Poly-Mount, (PolySciences). Mice were housed and procedures performed in accordance with European and national regulations (Protocol Authorisation APAFIS #38994-2023062718345638). Mice were bred in the CBI Specific and Opportunistic Pathogen Free (SOPF) animal facility. Mice were maintained on a C57Bl/6N genetic background.

**Data availability**

The data that support the findings of this study are available from the corresponding author upon request.

**Acknowledgements**

We acknowledge financial support from the Centre National de la Recherche Scientifique (CNRS), A*Midex (ANR-11-IDEX-0001-02, AMX-19-IET-002), ANR grants (ANR-10-INSB-04-01, ANR-11-INSB-0006, ANR-16-CONV-0001, ANR-21-ESRS-0002 IDEC), INSERM 22CP139-00, Chan Zuckerberg Initiative DAF (2024-337798), European Union's Horizon 2020 (EU ICT 101016923 CRIMSON and Marie Skłodowska-Curie Actions ITN 812992 MUSIQ) and European Research Council (ERCs, SpeckleCARS, 101052911 & sCiSsoRS, 101124764). Views and opinions expressed are however those of the author(s) only and do not necessarily reflect those of the European Union or the European Research Council. Neither the European Union nor the granting authority can be held responsible for them. The authors thank Annafrancesca Rigato for the preparation of the drosophila larva epidermis samples; Olivier Hector, Antonin Moreau and Julien Lumeau for the zero order stop custom mirror coating.

## Conflict of Interest

The authors declare no conflict of interest.

## Author contributions

A.B. prepared the samples, performed the experiments and developed the 2P-RIM app. G.G. and X.L. developed the 2P-RIM reconstruction algorithm. E.F., T.M., and F.V. assisted the samples preparation and experiments. A.S. conceived the idea, derived the analytical description of 2P-RIM, and supervised the project. H.R. conceived the idea and supervised the project. S.H. conceived the idea, assisted the experiments and supervised the project. A.B., X.L., H.R., A.S. and S.H. wrote the manuscript. All authors discussed the results and commented on the manuscript.

## References

1. W. Denk, J. H. Strickler, and W. W. Webb, "Two-photon laser scanning fluorescence microscopy," Science **248**, 73–76 (1990).
2. J. Wu, N. Ji, and K. K. Tsia, "Publisher correction: Speed scaling in multiphoton fluorescence microscopy," Nat. Photonics **16**, 87–87 (2021).
3. J. A. Feijó and N. Moreno, "Imaging plant cells by two-photon excitation," Protoplasma **223**, 1–32 (2004).
4. H. J. Koester, D. Baur, R. Uhl, and S. W. Hell, "Ca2+ fluorescence imaging with pico- and femtosecond two-photon excitation: Signal and photodamage," Biophys. J. **77**, 2226–2236 (1999).
5. A. Hopt and E. Neher, "Highly nonlinear photodamage in two-photon fluorescence microscopy," Biophys. J. **80**, 2029–2036 (2001).
6. F. Niu, R. Wu, D. Wu, D. Gou, L. Feng, L. Chen, Z. Zhang, and A. Wang, "Photobleaching of ultrashort pulses with different repetition rates in two-photon excitation microscopy," Laser Phys. **29**, 046001 (2019).
7. W. R. Zipfel, R. M. Williams, and W. W. Webb, "Nonlinear magic: multiphoton microscopy in the biosciences," Nat. Biotechnol. **21**, 1369–1377 (2003).
8. F. Helmchen and W. Denk, "Deep tissue two-photon microscopy," Nat. Methods **2**, 932–940 (2005).
9. M. Gu and C. Sheppard, "Comparison of three-dimensional imaging properties between two-photon and single-photon fluorescence microscopy," J. microscopy **177**, 128–137 (1995).
10. D. Oron and Y. Silberberg, "Harmonic generation with temporally focused ultrashort pulses," J. Opt. Soc. Am. B **22**, 2660 (2005).
11. G. Zhu, J. van Howe, M. Durst, W. Zipfel, and C. Xu, "Simultaneous spatial and temporal focusing of femtosecond pulses," Opt. Express **13**, 2153 (2005).
12. E. Papagiakoumou, E. Ronzitti, and V. Emiliani, "Scanless two-photon excitation with temporal focusing," Nat. Methods **17**, 571–581 (2020).
13. K. Isobe, T. Takeda, K. Mochizuki, Q. Song, A. Suda, F. Kannari, H. Kawano, A. Kumagai, A. Miyawaki, and K. Midorikawa, "Enhancement of lateral resolution and optical sectioning capability of two-photon fluorescence microscopy by combining temporal-focusing with structured illumination," Biomed. optics express **4**, 2396–2410 (2013).
14. A. Vaziri, J. Tang, H. Shroff, and C. V. Shank, "Multilayer three-dimensional super resolution imaging of thick biological samples," Proc. Natl. Acad. Sci. **105**, 20221–20226 (2008).
15. F. Cella Zanacchi, Z. Lavagnino, M. Faretta, L. Furia, and A. Diaspro, "Light-sheet confined super-resolution using two-photon photoactivation," PloS one **8**, e67667 (2013).
16. P. Szczypkowski, M. Pawlowska, and R. Lapkiewicz, "3d super-resolution optical fluctuation imaging with temporal focusing two-photon excitation," Biomed. Opt. Express **15**, 4381–4389 (2024).
17. T. Mangeat, S. Labouesse, M. Allain, A. Negash, E. Martin, A. Guénolé, R. Poincloux, C. Estibal, A. Bouissou, S. Cantaloube, E. Vega, T. Li, C. Rouvière, S. Allart, D. Keller, V. Debarnot, X. B. Wang, G. Michaux, M. Pinot, R. L. Borgne, S. Tournier, M. Suzanne, J. Idier, and A. Sentenac, "Super-resolved live-cell imaging using random illumination microscopy," Cell Reports Methods **1**, 100009 (2021).
18. K. Prakash, B. Diederich, R. Heintzmann, and L. Schermelleh, "Super-resolution microscopy: a brief history and new avenues," Philos. Transactions Royal Soc. A **380**, 20210110 (2022).
19. L. Mazzella, T. Mangeat, G. Giroussens, B. Rogez, H. Li, J. Creff, M. Saadaoui, C. Martins, R. Bouzignac, S. Labouesse *et al.*, "Extended-depth of field random illumination microscopy, edf-rim, provides super-resolved projective imaging," Light. Sci. & Appl. **13**, 285 (2024).

20. K. Affannoukoué, S. Labouesse, G. Maire, L. Gallais, J. Savatier, M. Allain, M. Rasedujjaman, L. Legoff, J. Idier, R. Poincloux *et al.*, "Super-resolved total internal reflection fluorescence microscopy using random illuminations," Optica **10**, 1009–1017 (2023).
21. E. M. Fantuzzi, S. Heuke, S. Labouesse, D. Gudavicius, R. Bartels, A. Sentenac, and H. Rigneault, "Wide-field coherent anti-stokes raman scattering microscopy using random illuminations," Nat. Photonics **17**, 1097–1104 (2023).
22. C. Ventalon and J. Mertz, "Quasi-confocal fluorescence sectioning with dynamic speckle illumination," Opt. Lett. **30**, 3350 (2005).
23. S. Labouesse, J. Idier, M. Allain, G. Giroussens, T. Mangeat, and A. Sentenac, "Superresolution capacity of variance-based stochastic fluorescence microscopy: From random illumination microscopy to superresolved optical fluctuation imaging," Phys. Rev. A **109**, 033525 (2024).
24. M. G. L. Gustafsson, "Surpassing the lateral resolution limit by a factor of two using structured illumination microscopy." J. Microsc. **198**, 82–87 (2000).
25. J. W. Goodman, *Speckle phenomena in optics: theory and applications* (Roberts and Company Publishers, 2007).
26. J. Mertz, *Introduction to optical microscopy* (Cambridge University Press, 2019).
27. G. Giroussens, S. Labouesse, M. Allain, T. Mangeat, L. Mazzella, L. le Goff, A. Sentenac, and J. Idier, "Fast super-resolved reconstructions in fluorescence random illumination microscopy (rim)," IEEE Transactions on Comput. Imaging (2024).
28. S. Wildman, A. M. Hirsch, S. Kirchanski, and D. Spencer, "Chloroplasts in living cells and the string-of-grana concept of chloroplast structure revisited," Discov. Photosynth. pp. 737–744 (2005).
29. S. Abrahamsson, J. Chen, B. Hajj, S. Stallinga, A. Y. Katsov, J. Wisniewski, G. Mizuguchi, P. Soule, F. Mueller, C. D. Darzacq *et al.*, "Fast multicolor 3d imaging using aberration-corrected multifocus microscopy," Nat. methods **10**, 60–63 (2013).
30. B. Zhao, M. Koyama, and J. Mertz, "High-resolution multi-z confocal microscopy with a diffractive optical element," Biomed. Opt. Express **14**, 3057–3071 (2023).
31. S. Labouesse, T. Mangeat, A. Sentenac, G. Giroussens, M. Allain, J. Idier, and C. Estibal, "Lightrim: Procédé d'obtention d'une image haute-résolution d'un échantillon," I-21-2566-02 Demande n 2309082 (2025).
32. F. Vernuccio, A. Benachir, E. M. Fantuzzi, B. Morel, S. Bux, E. Martin, J. Villanueva, Y. Pertot, N. Thiré, S. Heuke, and H. Rigneault, "Dual picosecond fast tunable optical parametric amplifier laser system for wide-field nonlinear optical microscopy," APL Photonics **9**, 096113 (2024).
33. S. Heuke, N. Vogler, T. Meyer, D. Akimov, F. Kluschke, H.-J. Röwert-Huber, J. Lademann, B. Dietzek, and J. Popp, "Multimodal mapping of human skin," Br. J. Dermatol. **169**, 794–803 (2013).

# Widefield two-photon random illumination microscopy (2P-RIM) - Supplementary information

**Assia Benachir[1], Xiangyi Li[1], Eric M. Fantuzzi[1], Guillaume Giroussens[1], Thomas Mangeat[2], Federico Vernuccio[1], Hervé Rigneault[1,*], Anne Sentenac [1,*], and Sandro Heuke[1,*]**

[1]*Aix Marseille Univ, CNRS, Centrale Med, Institut Fresnel, Marseille, France.*
[*]*Corresponding authors: anne.sentenac@fresnel.fr, sandro.heuke@fresnel.fr, herve.rigneault@fresnel.fr,*

**Abstract:** (1) We present the theory for the 2P-RIM algorithm and derive an expression for the preconditioned gradient that iteratively enhances the resolution of the standard deviation of a stack of speckle illuminated 2-photon fluorescence images. (2) We compare the speed or reduced photo-toxicity advantage of wide-field versus focused laser scanning 2P-fluorescence microscopy and consider two scenarios: first, we search for the speed advantage assuming to have the same peak power at the sample. Second, we search for the peak power difference assuming to have the same total acquisition time. (3) We compare the linear and nonlinear photo-toxicity in focused and wide-field 2P-excited fluorescence microscopy. (4) We introduce and describe a standalone app (free to download) for processing of series of speckle-illuminated 2P-excited fluorescence images. (5) We quantify 2P-RIM's axial and transverse spatial resolution on grounds of experimental data.

## 1. Modeling two-photon Random Illumination Microscopy: 2P-RIM

### 1.1. The image model

In two-photon RIM, one records multiple images of the same sample under different speckle illuminations (obtained for example by passing the pulsed laser beam through a rotating diffuser). We note $\boldsymbol{R}$ the transverse coordinate of a pixel of the camera space, hereafter noted $\Gamma$, and $\boldsymbol{r} + z\hat{\boldsymbol{z}}$, with $\hat{\boldsymbol{z}}$ the optical axis, a point in the object space, noted $\Omega$. If we assume that the microscope magnification is one, the intensity $I$ of the fluorescent light recorded on the camera can be modeled as

$$I(\boldsymbol{R}) = \int H(\boldsymbol{R} - \boldsymbol{r}, -z)\rho(\boldsymbol{r}, z)|E|^4(\boldsymbol{r}, z)\mathrm{d}\boldsymbol{r}\mathrm{d}z + b(\boldsymbol{R}) + \eta(\boldsymbol{R}) \quad (1)$$

where $\rho$ is the two-photon fluorescence density, $e$ is the complex electric field of the random illumination intensity (which is approximated by a scalar monochromatic wave), $b$ is a background light that is assumed to be independent of the illumination, $\eta$ is an uncorrelated zero-mean noise (gathering the Poisson and electronic noises), and $H$ is the microscope point spread function which depends on the numerical aperture of the objective NA and on the emission wavelength $\lambda$. The expression of $H$ is [1],

$$H(\boldsymbol{r}, z) = |\int p(\boldsymbol{\nu})e^{2i\pi[\boldsymbol{\nu}\cdot\boldsymbol{r}+\gamma(\boldsymbol{\nu})z]}\mathrm{d}\boldsymbol{\nu}|^2, \quad (2)$$

where $p$ is the coherent observation pupil function which is equal to 1 if $\nu < \mathrm{NA}/\lambda$ and 0 elsewhere, and $\gamma(\boldsymbol{\nu}) = \sqrt{1/\lambda^2 - \nu^2}$.

### 1.2. Modeling the speckled illumination and its statistics

We assume that the illumination is a fully developed speckle at the illumination wavelength $\lambda_{\mathrm{ex}}$ (which is roughly twice larger than the fluorescence wavelength $\lambda$). The speckle field can be written as a sum of plane waves with random phase,

$$E(r, z) = \int p_i(\nu) e^{i\phi(\nu)} e^{2i\pi[\nu \cdot r + \gamma_i(\nu)z]} d\nu, \quad (3)$$

with $\gamma_i(\nu) = \sqrt{1/\lambda_{\text{ex}}^2 - \nu^2}$, $\phi(\nu)$ is an uncorrelated random phase uniformly distributed between $[0, 2\pi]$, and $p_i$ is the coherent illumination pupil function which is generally similar to $p$ except that the fluorescence wavelength $\lambda$ is replaced by the illumination wavelength $\lambda_{\text{ex}}$.

In RIM, the reconstruction of the fluorescence density is obtained from the variance of images recorded under many realizations of the speckled illuminations. The reconstruction scheme requires the knowledge of the statistics of the illumination field. Since the field $E$ is a complex circular Gaussian random variable of zero mean (by virtue of the Central Limit Theorem), all its statistical moments depend on the field auto-covariance, $\mathbb{E}\{E(r, z)E^*(r', z')\} = C(r - r', z - z')$ where $\mathbb{E}\{\}$ indicates averaging over many realizations of the random speckled illumination. The autocovariance $C$ reads,

$$C(r, z) = \int |p|^2(\nu) e^{2i\pi(\nu \cdot r + \gamma(\nu)z)} d\nu. \quad (4)$$

The field intensity auto-covariance is

$$\mathbb{E}\{|E|^2(r, z)|E|^2(r', z')\} - \mathbb{E}\{|E|^2(r, z)\} \times \mathbb{E}\{|E|^2(r', z')\} = |C|^2(r - r', z - z') \quad (5)$$

and the field square intensity auto-covariance is [2],

$$\Lambda(r - r', z - z') = \mathbb{E}\{|E|^4(r, z)|E|^4(r', z')\} - \mathbb{E}\{|E|^4(r, z)\} \times \mathbb{E}\{|E|^4(r', z')\}$$
$$= 16C^2(0, 0)|C|^2(r - r', z - z') + 4|C|^4(r - r', z - z'). \quad (6)$$

### 1.3. Modeling the variance of images obtained under random speckled illuminations

Hereafter, we assume that the sample $\rho$ is a thin slice located at the object's focal plane $\rho(r, z) = \rho(r)\delta(z)$. We assume that the signal stemming from the out-of-focus region is correctly modeled by the background noise $b$. In this case, the image model becomes,

$$I(R) = \int H(R - r)\rho(r)|E|^4(r) dr + b(R) + \eta(R), \quad (7)$$

where $H(R - r) = H(R - r, 0)$ and $E(r) = E(r, 0)$. Using this simplification, the camera space $\Gamma$ becomes equivalent to the object space $\Omega$. However, we will still use the notation $R$ for a point of $\Gamma$ and $r$ for a point of $\Omega$ to keep a slight distinction between the two spaces.

In RIM, one forms the variance of multiple speckled images obtained under different realizations of the speckle field, $\sigma^2(R) = \mathbb{E}\{I^2(R)\} - \mathbb{E}\{I(R)\}^2$. Using the knowledge of the speckle fields moments Eq. (6), the expression of the variance reads,

$$\sigma^2(R) = \int H(R - r_1)\rho(r_1)\Lambda(r_1 - r_2)H(R - r_2)\rho(r_2) dr_1 dr_2 + v_\eta \quad (8)$$

$$= \int T(R - r_1, R - r_2)\rho(r_1)\rho(r_2) dr_1 dr_2 + v_\eta(R), \quad (9)$$

with $T(x, y) = H(x)\Lambda(x - y)H(y)$ and $v_\eta$ is the variance of the noise *which is assumed to be known* (the estimation of $v_\eta$ is detailed in the practical implementation of the inversion scheme in section 4). RIM consists in recovering $\rho$ from the variance of the speckled images. It has been shown in [3] that if the support of $\Lambda$ is the same as that of $H$, the variance uniquely defines the spatial frequencies of the sample $\rho$ in the Fourier support of $H^2$ (which is significantly larger than that of $H$). Now, the Fourier support of $H$ is a disk of radius $2NA/\lambda$ while that of $C$ is a disk of radius $NA/\lambda_{\text{ex}}$. In two-photon microscopy, $\lambda_{\text{ex}} \approx \lambda/2$, so that the Fourier support of $C^4$ (and consequently $\Lambda$) is close to that of $H$. Two-photon RIM is thus, in principle, able to recover the sample spatial frequencies on the Fourier support of $H^2$, which corresponds to that of an ideal one-photon confocal microscope.

However, the reconstruction procedure is not trivial as the variance depends quadratically on the sample. Hence, we developed an iterative reconstruction procedure, algoRIM, that estimates the sample so as to minimize a distance between the theoretical expected variance and the experimental one. To be numerically tractable, the inversion scheme requires the fast calculation of the variance, Eq. (9), for a given sample estimate [4]. To this aim, we note that the integral operator $T$ in Eq. (9) is positive definite (i.e. it is a symmetric real matrix). It is thus possible to decompose $T$ as,

$$T(\boldsymbol{x}, \boldsymbol{y}) = \sum_{k=1}^{\infty} Q_k(\boldsymbol{x}) Q_k(\boldsymbol{y}) \tag{10}$$

where $Q_k$ are real orthogonal eigenfunctions in the sense of the scalar product

$$\langle f, g \rangle = \int f(\boldsymbol{r}) g(\boldsymbol{r}) \mathrm{d}\boldsymbol{r}, \tag{11}$$

of decreasing norm.

Usually, taking $K = 10$ eigenvectors is sufficient to get an accurate estimation of the variance. The calculation of the latter can then be done rapidly with few convolutions,

$$\sigma^2(\boldsymbol{R}) = \sum_{k=1}^{K} \left( \int Q_k(\boldsymbol{R} - \boldsymbol{r}_1) \rho(\boldsymbol{r}_1) \mathrm{d}\boldsymbol{r}_1 \right)^2 + v_\eta(\boldsymbol{R}). \tag{12}$$

which is rewritten in shorter notations as,

$$\sigma^2 = \sum_{k=1}^{K} (Q_k \otimes \rho)^2 + v_\eta. \tag{13}$$

where $\otimes$ stands for the convolution operator and the dependence on $\boldsymbol{R}$ has been omitted.

### 1.4. Recovering the fluorescence density from the variance of speckled images: a linear deconvolution

We now assume that we have measured an experimental variance of the speckled images which is noted $\hat{\sigma}^2$. We also assume that the noise variance is negligible (or is known and has been removed). The first technique for recovering $\rho$ from $\hat{\sigma}^2$ is based on the observation that the norm of the eigenvectors decays quickly when $k$ increases. Thus, a rough estimate of the theoretical variance is given by the first-order approximation of Eq. (12), $\sigma^2 \approx (Q_1 \otimes \rho)^2$.

The square root of $\sigma^2$, i.e. the theoretical standard deviation of the speckled images $\sigma$, is then linearly linked to the fluorescence density through a convolution operator,

$$\sigma \approx Q_1 \otimes \rho. \tag{14}$$

It is thus possible to recover an estimation of the fluorescence density from the experimental data using a simple Tikhonov filter $f$,

$$\rho^{\text{est}} \approx \hat{\sigma} \otimes f \tag{15}$$

with

$$\tilde{f}(\boldsymbol{\nu}) = \frac{\tilde{Q}_1^*(\boldsymbol{\nu})}{|\tilde{Q}_1|^2(\boldsymbol{\nu}) + \mu} \tag{16}$$

where $\tilde{f}$ indicates the 2D-fourier transform of $f$, $\tilde{f}(\nu) = \int f(r)\exp(-2i\pi\nu\cdot r)\mathrm{d}r$ and $\mu$ is a scalar that is manually tuned to reduce the influence of noise.

## 1.5. Recovering the fluorescence density from the variance of speckled images: Iterative inversion scheme

To improve further the estimation of the fluorescence density, we also developed an iterative inversion scheme avoiding the first-order approximation of the standard deviation. In general, our model for the standard deviation of the speckled images can be seen as applying a non-linear operator to $\rho$,

$$\sigma = \mathcal{S}[\rho] = \sqrt{\sum_1^K (Q_k \otimes \rho)^2 + v_\eta}. \tag{17}$$

Hereafter the notation $\mathcal{W}[f]$ indicates the result (either a scalar or a function) of the functional $\mathcal{W}$ (a functional is an operator acting on functions) applied to the function $f$.

Our inversion scheme consists in finding $\rho$ that minimizes the cost functional $\mathcal{F}$,

$$\mathcal{F}[\rho] = \|\hat{\sigma} - \mathcal{S}[\rho]\|^2 + \mu\|\rho\|^2 \tag{18}$$

where $\|f\|^2 = \langle f, f \rangle$ as defined in Eq. (11), and the term $\mu\|\rho\|^2$ is a Tikhonov regularization.

The chosen minimization procedure, which is based on a gradient descent, requires the linearization of the data model.

### 1.5.1. Differentiating the standard deviation operator

We first introduce the differential (Jacobian for discrete problems) $\mathcal{J}_\rho$ of the functional $\mathcal{S}$. It is the operator acting on a function $u$ on $\Omega$ and forming a function on $\Gamma$ defined as,

$$\mathcal{J}_\rho[u] = \lim_{t\to 0} \frac{\mathcal{S}[\rho + tu] - \mathcal{S}[\rho]}{t},$$
$$\mathcal{J}_\rho[u] = \frac{\sum_1^K (Q_k \otimes \rho) \times (Q_k \otimes u)}{\sigma}. \tag{19}$$

We observe that $\mathcal{J}_\rho[u]$ is linear with respect to $u$.

We also define the second-order differential of $\mathcal{S}$ as the operator $\mathcal{D}\mathcal{J}_\rho$ acting on functions $(u, v)$ on $\Omega$ and forming a function on $\Gamma$ such that,

$$\mathcal{D}\mathcal{J}_\rho[u, v] = \lim_{t\to 0} \frac{\mathcal{J}_{\rho+tu}[v] - \mathcal{J}_\rho[v]}{t},$$
$$\mathcal{D}\mathcal{J}_\rho[u, v] = \frac{\sum_1^K (Q_k \otimes u) \times (Q_k \otimes v)}{\sigma} - \frac{\mathcal{J}_\rho[u] \times \mathcal{J}_\rho[v]}{\sigma}. \tag{20}$$

We observe that $\mathcal{D}\mathcal{J}_\rho[u, v]$ is linear with respect to $u$ and $v$. Hereafter, we note $\mathcal{A}$ the linear operator that depends on $(u, \rho)$ such that $\mathcal{D}\mathcal{J}_\rho[u, v] = \mathcal{A}[v]$.

### 1.5.2. Inversion procedure

In this section, we describe the inversion procedure which is based on a classical preconditioned gradient descent. Even though most of the notions described here are well known, in particular in the signal processing community, we thought it useful to present them in detail for the optical community.

To minimize iteratively the cost functional, we start from an estimate of the fluorescence density $\rho$. We calculate the residue

$$h_\rho = \hat{\sigma} - \mathcal{S}[\rho] \tag{21}$$

and form the cost functional
$$\mathcal{F}[\rho] = \|h_\rho\|^2 + \mu\|\rho\|^2. \tag{22}$$

Then, we modify $\rho$ along a direction $u$ (we recall that $u$ is a function on $\Omega$) by forming $\rho^{\text{new}} = \rho + \alpha u$ where $\alpha$ is a scalar and we calculate $\alpha$ so that the scalar function $l(\alpha) = \mathcal{F}[\rho + \alpha u]$ is minimal. The main difficulty is to find an efficient search direction $u$. To this aim, we introduce the differential of the cost functional $\mathcal{DF}_\rho$ which is the linear operator defined as,

$$\begin{aligned} \mathcal{DF}_\rho[u] &= \lim_{t \to 0} \frac{\mathcal{F}[\rho + tu] - \mathcal{F}[\rho]}{t}, \\ &= -2\langle h_\rho, \mathcal{J}_\rho[u]\rangle + 2\langle \mu\rho, u\rangle. \end{aligned} \tag{23}$$

To pursue our analysis, we have to introduce the notion of adjoint operator. The adjoint of any linear operator $\mathcal{W}$, is the linear operator, noted $\mathcal{W}^\dagger$, that satisfies,

$$\langle u, \mathcal{W}[v]\rangle_\Gamma = \langle \mathcal{W}^\dagger[u], v\rangle_\Omega. \tag{24}$$

If $\mathcal{W}$ is an integral operator such that $\mathcal{W}[v] = \int_\Omega \mathcal{W}(\mathbf{R}, \mathbf{r})v(\mathbf{r})d\mathbf{r}$ then, the adjoint operator is also a linear integral operator such that, $\mathcal{W}^\dagger[u] = \int_\Gamma \mathcal{W}(\mathbf{R}, \mathbf{r})u(\mathbf{R})d\mathbf{R}$. Note that in our configuration $\Omega = \Gamma = \mathbb{R}^2$.

Using the definition of the adjoint, Eq. (23) can be rewritten as,

$$\mathcal{DF}_\rho[u] = 2\langle -\mathcal{J}_\rho^\dagger[h_\rho] + \mu\rho, u\rangle. \tag{25}$$

Using the Cauchy-Schwarz inequality, $\langle f, u\rangle^2 \leq \langle f, f\rangle\langle u, u\rangle$, one deduces that the function $u$ of norm unity, $\|u\|^2 = \langle u, u\rangle = 1$, that maximizes $|\mathcal{DF}_\rho[u]|$ is given by $u = (-\mathcal{J}_\rho^\dagger[h_\rho] + \mu\rho)/\|-\mathcal{J}_\rho^\dagger[h_\rho] + \mu\rho\|$. Hereafter we introduce $g_\rho$ the function on $\Omega$,

$$g_\rho = -\mathcal{J}_\rho^\dagger[h_\rho] + \mu\rho, \tag{26}$$

which is called the gradient of the cost-functional. The gradient indicates the direction that maximizes the local variation of the cost-functional. In a standard steepest descent minimization procedure, the searching direction is chosen equal to $g_\rho$. Indeed, this direction ensures that the cost-functional will be strongly modified at each iteration.

The expression of $g_\rho$ from Eq. (26,19) is,

$$g_\rho(\mathbf{r}) = -\sum_1^K \int d\mathbf{R} Q_k(\mathbf{R} - \mathbf{r})\frac{h_\rho(\mathbf{R})}{\sigma(\mathbf{R})} \int Q_k(\mathbf{R} - \mathbf{r}_1)\rho(\mathbf{r}_1)d\mathbf{r}_1 + \mu\rho(\mathbf{r}). \tag{27}$$

It is easily calculated in the Fourier space, the Fourier transform, $\tilde{g}_\rho$ of the gradient being,

$$\tilde{g}_\rho(\nu) = -\sum_1^K \tilde{Q}_k^*(\nu) \int \tilde{Q}_k(\nu - \nu')\tilde{\rho}(\nu - \nu')\tilde{s}(\nu')d\nu' + \mu\tilde{\rho}(\nu), \tag{28}$$

where $s = h_\rho/\sigma$.

However, we have observed that the inversion procedure using the gradient as the search direction usually requires too many iterations to converge. We have thus looked for another descent direction to fasten the inversion process.

Ideally, the fluorescence density estimate should be modified by $u$ so that the differential of the cost functional taken at $\rho + u$ applied to any vector $v$, $\mathcal{DF}_{\rho+u}[v]$, is equal to 0. Indeed, extrema or saddle points of the cost-functional are characterized by the nullity of the differential at these points. Hereafter, we assume that our first estimate $\rho$ is not far from an extremum so that $\|u\|$ is small. The differential of the cost-functional at $\rho + u$ reads,

$$\mathcal{DF}_{\rho+u}[v] = -2(\langle -g_\rho - \mu u, v \rangle) - \langle \mathcal{J}_\rho[u], \mathcal{J}_\rho[v] \rangle + \langle h_\rho, \mathcal{DJ}[u,v] \rangle) + o(\|u\|^2). \quad (29)$$

Recalling that $\mathcal{DJ}[u,v] = \mathcal{A}[v]$ and introducing the adjoints, Eq. (29) can be cast as,

$$\mathcal{DF}_{\rho+u}[v] = -2(\langle -g_\rho - \mu u - \mathcal{J}_\rho^\dagger[\mathcal{J}_\rho[u]] + \mathcal{A}^\dagger[h_\rho], v \rangle) + o(\|u\|^2). \quad (30)$$

Then, the condition $\mathcal{DF}_{\rho+u}[v] = 0$ whatever $v$ implies that,

$$-g_\rho - \mu u - \mathcal{J}_\rho^\dagger[\mathcal{J}_\rho[u]] + \mathcal{A}^\dagger[h_\rho] = 0. \quad (31)$$

Using Eq. (20) and the property $\langle h_\rho, f \times \mathcal{W}[v] \rangle_\Gamma = \langle \mathcal{W}^\dagger[h_\rho \times f], v \rangle_\Omega$, we show that,

$$\mathcal{A}^\dagger[h_\rho] = \sum_1^K Q_k^\dagger \left[ \frac{h_\rho}{\sigma} \times (Q_k \otimes u) \right] - \mathcal{J}_\rho^\dagger \left[ \frac{h_\rho}{\sigma} \times \mathcal{J}_\rho[u] \right]. \quad (32)$$

The explicit formulation of Eq. (31) is then,

$$-\mu u - \sum_1^K Q_k^\dagger [Q_k \otimes u] + \sum_1^K Q_k^\dagger \left[ \frac{\hat{\sigma}}{\sigma} \times (Q_k \otimes u) \right] - \mathcal{J}_\rho^\dagger \left[ \frac{\hat{\sigma}}{\sigma} \times \mathcal{J}_\rho[u] \right] = g_\rho. \quad (33)$$

where we have used the equality $\frac{h_\rho}{\sigma} = \frac{\hat{\sigma}}{\sigma} - 1$.

Finding $u$, solution of the linear system Eq. (33) is computationally demanding and in practice impossible. However, if the closed form approximation is valid, i.e. if one retains only the first term in the eigenvector expansion so that, $\sigma \approx Q_1 \otimes \rho$ and $\mathcal{J}_\rho[u] \approx Q_1 \otimes u$, the two last terms of the left-handed part of Eq. (33) cancel each other and the linear system simplifies in,

$$-\mu u - Q_1^\dagger [Q_1 \otimes u] \approx g_\rho \quad (34)$$

Recalling that $Q_1^\dagger[Q_1 \otimes u](r) = \int Q_1(R-r) Q_1(R-r') u(r') dR dr'$, we can rewrite equation (34) in the Fourier space as,

$$-(|\tilde{Q}_1|^2(\nu) + \mu) \tilde{u}(\nu) \approx \tilde{g}_\rho(\nu). \quad (35)$$

Under the first order approximation, the gradient, Eq. (27), simplifies in $\tilde{g}_\rho(\nu) \approx -\tilde{Q}_1^*(\nu) \tilde{h}_\rho(\nu)$. Thus, in one step, we recover the same expression as the Tikhonov deconvolution Eq. (15,16) for the estimation of the fluorescence density.

Even if we account for several terms in the eigenvector expansion of $\sigma$, it is likely that the difference between the two last terms of the left-handed part of Eq. (33) remains smaller than the sum of the first two terms. Then, an approximate direction $u$ can be obtained by solving the simplified linear system,

$$-\mu u - \sum_1^K Q_k^\dagger [Q_k \otimes u] \approx g_\rho. \quad (36)$$

The expression of the Fourier transform of $u$ is then,

$$\tilde{u}(\nu) = -\frac{\tilde{g}_\rho(\nu)}{\sum_1^K |\tilde{Q}_k(\nu)|^2 + \mu}. \quad (37)$$

This novel searching direction can be seen as a preconditioned gradient. Its estimation at each iteration is not more time-consuming than the gradient, Eq. (37), since $\sum_1^K |\tilde{Q}_k(\nu)|^2 + \mu$ does not depend on the estimate $\rho$ and can be calculated only once at the beginning of the iterations. We observed that this searching direction allowed a significant acceleration of the convergence of the algorithm.

*1.6. Implementation of the inversion procedure*

In practice, one records $P$ fluorescence images $I_p^{\text{raw}}$, $p = 1...P$, for $P$ different realizations of the speckled illumination. We assume that the point spread function of the microscope is known and we call it $H_{\text{raw}}$. We also assume that the illumination pupil function $p_i$ used for forming the speckled illumination, Eq. (3), is known.

The recorded signal is discretized over the pixels of the camera, so the maximum frequency of the images is $1/d$ where $d$ is the camera pixel size after demagnification. Generally, $d$ is chosen following the Nyquist criterium, $d \approx \lambda/(4\text{N}A)$, to maximize the number of photons per pixel without losing high-frequency information.

The first data processing consists in increasing artificially the maximum spatial frequency of the images (via zero-padding) to get $d^{\text{new}} \leqslant \lambda/(8\text{N}A)$. This interpolation is mandatory for estimating properly the variance of the speckled images and getting a super-resolved reconstruction.

The second step is the estimation of the noise and speckled images' variances.

*1.6.1. Data preprocessing*

Estimation of the noise variance:

From Eq. (1) we observe that the spectrum of the fluorescence signal in $I_p$ belongs necessarily to the Fourier support $D$ of the microscope point spread function, $H_{\text{raw}}$. Ideally, $D$ is a disk of radius $2\text{N}A/\lambda$. On the other hand, the spectrum of the noise covers the whole Fourier space. To keep only the noise information, we crop the images' frequencies by a filter $\tilde{q}$ such that $\tilde{q}(\nu) = 0$ for $\nu \in D$ and one elsewhere. We assume that the resulting images correspond to the pure noise information convolved with $q$. At this point, we recall that if $\eta$ is an uncorrelated random noise of variance $\nu_\eta$, the variance of $\eta \otimes q$ is equal to $\nu_\eta \otimes q^2$. The (convolved) noise variance is estimated empirically through,

$$\nu_{\text{conv}}(\boldsymbol{R}) = \frac{\sum_{p=1}^P J_p^2(\boldsymbol{R})}{P} - \frac{[\sum_{p=1}^P J_p(\boldsymbol{R})]^2}{P^2}, \quad (38)$$

where $J_p = I_p^{\text{raw}} \otimes q$. Now, if the maximum spatial frequency $1/d$ is larger than $4\text{N}A/\lambda$, the Fourier transform of $q^2$, $\tilde{q} \otimes \tilde{q}$, is never equal to 0. Thus, a simple deconvolution can be used to recover $\nu_\eta$ from $\nu_{\text{conv}}$: $\tilde{\nu}_\eta = \tilde{\nu}_{\text{conv}}/(\tilde{q} \otimes \tilde{q})$.

Estimation of the variance of the speckled images:

We now turn to the calculation of the empirical variance of the speckled images, $\hat{\sigma}^2$. It is not directly calculated from the raw images $I_p^{\text{raw}}$. Indeed, it is possible to improve the data by removing the noise beyond $D$ and enhancing the high frequencies damped by $H_{\text{raw}}$. To this aim, the images $I_p^{\text{raw}}$ are prefiltered with a Wiener filter $f$,

$$I_p = I_p^{\text{raw}} \otimes f, \quad (39)$$

with

$$\tilde{f} = \frac{\tilde{H}_{\text{raw}}^*}{|\tilde{H}_{\text{raw}}|^2 + \tau}, \quad (40)$$

where $\tau$ is a Tikhonov parameter that needs to be tuned manually. Then, the experimental variance at pixel $\boldsymbol{R}$ of the camera is estimated with,

$$\hat{\sigma}^2(\boldsymbol{R}) = \frac{\sum_{p=1}^P I_p^2(\boldsymbol{R})}{P} - \frac{[\sum_{p=1}^P I_p(\boldsymbol{R})]^2}{P^2}. \quad (41)$$

The convolution of the raw images by $f$ modifies the point spread function and the noise. The point spread function appearing in Eq. (1) is given by $H = H_{\text{raw}} \otimes f$ while the noise variance is estimated through, $\tilde{v}_\eta = \tilde{v}_{\text{conv}} \times \frac{\tilde{f} \otimes \tilde{f}}{\tilde{q} \otimes \tilde{q}}$. On the other hand, the speckle correlation $\Lambda$ is not modified by the image convolution. Once $H$ and $\Lambda$ are known, we calculate the eigenvectors $Q_k$, $k = 1..K$ (with $K \leq 10$ in general) from Eq. (10) to ensure a quick calculation of the theoretical standard deviation.

1.6.2. Iterative inversion

A closed-form estimation of the fluorescence density is found using (15,16). Note that this procedure requires adjusting a second Tikhonov parameter manually, $\mu$.

In the iterative inversion scheme, we start with an initial guess for the fluorescence density $\rho_0$ (either a constant, the closed-form result, or the experimental standard deviation) and an estimation of the Tikhonov parameter $\mu$.

At the $m^{\text{th}}$ iteration ($m > 0$), we calculate the preconditioned gradient direction $u_m$ with Eq. (37) using the estimation of the fluorescence density at the previous step, $\rho_{m-1}$.

We improve the descent direction through the Polak-Ribière gradient combination. For $m = 1$, the descent direction $d_1$ is equal to $u_1$ and for $m > 1$,

$$d_m = u_m + \beta_m d_{m-1}, \tag{42}$$

with

$$\beta_m = \frac{\langle u_m, u_m - u_{m-1} \rangle}{\|u_{m-1}\|^2}. \tag{43}$$

We modify the fluorescence density following, $\rho_m = \rho_{m-1} + \alpha d_m$. The scalar $\alpha$ is ideally chosen so as to minimize $l_m(\alpha) = \mathcal{F}[\rho_{m-1} + \alpha d_m]$. In practice, we use the first Newton step, $\alpha = l'_m(0)/l''_m(0)$,

$$\alpha = \frac{-\langle h_\rho, \mathcal{J}[d_m] \rangle + \mu \langle \rho_{m-1}, d_m \rangle}{\|\mathcal{J}[d_m]\|^2 - 2\langle h_\rho, \mathcal{DJ}[d_m, d_m] \rangle + \mu \|d_m\|^2}, \tag{44}$$

which is easily calculated using Eqs. (19,20).

The iterations are continued until $\mathcal{F}[\rho]$ gets smaller than a given value or for a predefined maximum number of iterations.

We have observed that the inversion scheme converged towards similar solutions whatever the initial guess. On the other hand, the quality of the reconstruction depended on the tuning of the two Tikhonov parameters, $\tau$ and $\mu$. As a rule of thumb, $\tau$ should be taken the smallest possible to give a chance to the inversion procedure to recover the high spatial frequencies of the sample hidden in the high spatial frequencies of the speckled images.

## 2. Signal strength: focused laser scanning versus widefield two-photon (2P) excited fluorescence microscopy

In this section, we discuss the ability of widefield 2P excited fluorescence to decrease the energy surface density inside the sample or to accelerate the image acquisition compared to the 2P-fluorescence point scanning scheme. We consider a simple imaging experiment in which the sample is 'thin' (2D imaging) and homogeneous. We assume that the excitation beam illuminates a small area $a$ in the point scanning focused case, or a much larger area $A$ in the wide-field case. The power per pulse is given by

$$I^u = P^u / N^u_{\text{rep}} \tag{45}$$

where $u = A$ stands for the widefield configuration while $u = a$ stands for the scanning configuration. $P^u$ is the laser power (in Watt) and $N^u_{\text{rep}}$ (in Hz) is the number of laser pulses per second. We further assume that the pulse duration is the same in the wide-field and point scanning cases which is why it does not appear as a variable within the equations below. The peak of energy surface density per pulse is given in the focused or widefield case by $I^u/u$, $u = (a, A)$. The fluorescence signal obtained for one laser pulse is proportional to the square of the excitation intensity integrated over the surface of the sample:

$$c^u \propto (I^u)^2/u \quad \text{with} \quad u = (a, A). \tag{46}$$

We now consider the global fluorescence signal that is obtained at the end of the image acquisition time. The fluorescence signal that is obtained when scanning the sample with a focused laser beam is

$$C^a = c^a \times N^a \times M, \tag{47}$$

where $N^a$ is the number of pulses used at one focal point and $M$ is the number of focal positions required to scan the whole sample. On the other hand, the 2P-fluorescence signal that is obtained for the whole image in the wide-field case is $C^A = c^A N^A$ where $N^A$ is the number of pulses used to get one widefield image.

The acquisition time for the case of laser scanning is

$$\Delta t^a = N^a \times M/N^a_{\text{rep}}. \tag{48}$$

Note that the number of pulses per pixel $N^a/N^a_{\text{rep}}$ cannot be diminished below the time required by the scanning system for moving from one focus position to the other which is a technical limitation of galvo-driven laser scanning systems. As a second speed limitation, at least one laser pulse per pixel is required. The acquisition time for the widefield image is

$$\Delta t^A = N^A/N^A_{\text{rep}} \tag{49}$$

Now it is possible to define the power $P^A$ and repetition rate $N^A_{\text{rep}}$ of the wide-field laser so that the wide-field image acquisition time is smaller than that of the scanning configuration while keeping the same fluorescence signal and peak of energy density. We introduce the factor $\alpha$ ($\alpha < 1$) that quantifies the ratio between the wide-field and point scanning acquisition times $\Delta t^A = \alpha \Delta t^a$. Under the constraints $C^a = C^A$ (same signal) and $I^a/a = I^A/A$ (same peak power), we find that

$$P^A = P^a/\alpha \tag{50}$$

and

$$N^A_{\text{rep}} = N^a_{\text{rep}} \frac{a}{\alpha A}. \tag{51}$$

For example, if $a/A = 1.25 \times 10^{-5}$ (typically an image of $A = 200 \times 200$ $\mu m^2$ with a scanning spot of $a = 0.5$ $\mu m^2$) and a scanning laser power $P^a = 10$ mW at a repetition rate $N^a_{\text{rep}} = 80$ MHz, we can diminish the acquisition time by hundred, $\alpha = 0.01$, with $P^A = 1$ W and $N^A_{\text{rep}} = 100$ kHz. Alternatively, we can also define the power and repetition rate of the wide-field laser for the peak of energy surface density in the wide-field scheme to be $\alpha$ times that of the point scanning scheme ($\alpha I^a/a = I^A/A$) while keeping the same image acquisition time ($\Delta t^a = \Delta t^A$) and same fluorescence signal ($C^a = C^A$). In this case, we obtain,

$$P^A = P^a/\alpha \tag{52}$$

and

$$N^A_{\text{rep}} = N^a_{\text{rep}} \frac{a}{\alpha^2 A}. \tag{53}$$

If we keep the same scanning parameters as previously considered ($A = 200 \times 200$ µm$^2$, $a = 0.5$ µm$^2$, $P^a = 10$ mW, $N^a_{\text{rep}} = 80$ MHz and $N^A_{\text{rep}} = 100$ kHz) we can diminish the peak power by a factor of ten, $\alpha = 0.1$, with $P^A = 1$ W which corresponds roughly to our wide field experimental parameters.

This simple analysis shows how to design laser parameters (power and repetition rate) in the wide-field scheme to improve 2P excited fluorescence imaging, either by reducing the peak energy in the sample or by increasing the image acquisition speed. Controlling independently the peak energy and the acquisition speed is just not possible in the point scanning scheme.

Finally, we note that we assumed a homogeneous illumination of the sample in this simple derivation while our experimental illumination is a speckle. In the speckle case, the excitation field has the same average power but is locally either enhanced (bright speckle grain) or reduced (dark speckle grain). However, the additional fluorescence signal from bright grains more than compensates for the dark speckle grains. In fact, the speckle illumination may be considered as a densely packed multi-focus pattern with random focus positions.

### 3. Phototoxicity: Linear and non-linear sample damage considerations

In nonlinear microscopy, the sample can be damaged by linear absorption of the laser followed by heating or as a result of multi-photon absorption leading to disruption of molecular bonds or ionization [5]. In the following, we investigate the linear and non-linear photo-damage depending on the illumination scheme (wide-field or point scanning). To simplify the comparison, we approximate our speckle fields by a homogeneous wide-field illumination and assume that the speckle intensity oscillates in time and space about the average wide-field intensity. Non-linear photo-damage is induced by high field strength and requires the comparison of the amount of high-order multi-photon absorption events ($D^u$). The latter is given for wide-field (A) and focused microscopy (a) by the product of the illuminated area, number of laser pulses and laser intensity:

$$D^u = u N_{\text{rep}}^u \left(\frac{I^u}{u}\right)^O \quad (54)$$

where $O$ is the order of non-linearity. For example, $O = 3$ corresponds to 3-photon absorption processes. We consider now the case where the image acquisition time ($\Delta t^a = \Delta t^A$) and the amount of 2P fluorescence signal ($C^a = C^A$) are the same under wide-field and scanning configuration. Within the previous section, we found (Eq. 52) that for these conditions $P^A = P^a/\alpha$. Using Eqs. 45 and 54, the ratio of the amounts of non-linear damage in scanning and wide-field microscopy can be derived as:

$$\frac{D^a}{D^A} = \left(\frac{N_{\text{rep}}^A A}{N_{\text{rep}}^a a}\right)^{O-1} \alpha^O \quad (55)$$

Using parameters that are close to our experimental conditions ($A = 200 \times 200$ µm$^2$, $a = 0.5$ µm$^2$, $N_{\text{rep}}^A$=100 kHz, $N_{\text{rep}}^a$=80MHz and $\alpha$=0.1), non-linear damage ratios $D^a/D^A$ of 1, 10, 100 and 1000 can be computed for the order of non-linearity ($O$) of 2, 3, 4 and 5, respectively. This means that wide-field 2P-excited fluorescence always generates less or equal chemical bond-breaking multi-photon absorption events. However, as soon as the dominant damage mechanism is equal or larger than order 3 ($O \geqslant 3$) much less photodamage is induced in the widefield case. Under NIR excitation, this condition is fulfilled for the majority of biomedical samples, because breaking of covalent chemical single bonds requires an energy larger than 5 $eV$, corresponding to at least 3 photons for $\lambda_{ex} \leq 750$nm. Thus, the laser pulses in wide-field 2P-excited fluorescence are less likely to trigger high-order multi-photon absorption events and features, therefore, lower non-linear phototoxicity.

Let us now consider the heating issue, where a minor fraction of every laser pulse is absorbed by the sample and leads to a rise in temperature. In point scanning microscopy, the absorbed energy ($Q^a_{\text{abs}}$) is transformed into heat ($Q^a_{\text{heat}}$) following,

$$\begin{aligned} Q^a_{\text{heat}} &= Q^a_{\text{abs}} \\ c_p \rho_d V \Delta T^a &= \sigma_{\text{abs}} l P^a \Delta t^a_{\text{abs}} \\ \Delta T^a &= \frac{\sigma_{\text{abs}} P^a \Delta t^a_{\text{abs}}}{c_p \rho_d a}, \end{aligned} \quad (56)$$

where $c_p$ is the specific heat capacity, $\rho_d$ is the mass density, $V = al$ ($l$ is the thickness of the sample), is the absorbing sample volume and $\Delta T^a$ is the local temperature rise. The right side of Eq. 56 is defined by the product of the absorption cross section $\sigma_{\text{abs}}$ with the laser power $P^a$ and the time span of illumination $\Delta t^a_{\text{abs}}$ at a given focal position. In scanning microscopy, the sample is illuminated, and therefore heated, locally over a short time span. The temperature rises to a maximum and the cooling starts as soon as the laser focus position is changed.

In nonlinear wide-field microscopy, the laser power per area is lower than in laser scanning 2P microscopy but the illumination time $\Delta t^A_{\text{abs}}$ is much longer, stacking the heat coming from a large number of pulses, but also allowing for an effective cooling of the sample. Therefore, in the wide-field case, the temperature of the sample increases with every laser pulse until an equilibrium between the laser heating and the cooling is established. Considering heat conduction as the dominant cooling mechanism, this equilibrium is reached for:

$$\begin{aligned} Q^A_{\text{con}} &= Q^A_{\text{abs}} \\ k \frac{\Delta T^A}{l} A \Delta t^A_{\text{abs}} &= \sigma_{\text{abs}} l P^A \Delta t^A_{\text{abs}} \\ \Delta T^A &= \frac{\sigma_{\text{abs}} l^2 P^A}{kA} \end{aligned} \quad (57)$$

Where $k$ is the conductivity constant and $\Delta T/l$ is the temperature gradient. To compare the heating in laser scanning and wide-field microscopy, we select parameters that correspond to the approximately 20µm thin U2O2 cells as presented in the main manuscript. We assume that the tissue is placed between an object holder and a coverslip made of fully transparent glass that both act as a heat sink of infinite capacity. The specific heat capacity $c_p$ and mass density $\rho_d$ are approximated by the corresponding values of water. The illuminated surfaces are $a$=0.5µm² and $A$=200×200 µm², the incident power $P^a$=10mW and $P^A$=1W, the absorption cross-section is $\sigma_{abs}$=1mm$^{-1}$ at 1030nm [6] and the conductivity constant of tissue is approximated by $k = 0.5 W/m/K$ [7]. The distance of the heated tissue to the heat sink (object holder) is at maximum half the size of the sample ($l$=10 µm) because there are heat sinks on both sides. We then set that the scanning image is acquired within the same time as the wide-field image ($\Delta t^A = \Delta t^a$). We set $\Delta t^A$=50 ms (20Hz) for the image acquisition time as in our experiments. The time during which the focused spot illuminates the same area is given by:

$$\Delta t^a_{\text{abs}} = \Delta t^A a / A \quad (58)$$

Entering the above-mentioned parameters into Eqs. 56-58 results in a short time temperature increase of $\Delta T^a$=12K in scanning microscopy and an equilibrium temperature of only $\Delta T^A$=5K above the temperature of the object holder for widefield 2P-excited fluorescence.

In summary, our wide-field illumination scheme provides an approximately 10 times lower peak energy density than point scanning (for the same fluorescence signal and acquisition time). This leads to at least a 100 times less multi-photon absorption events responsible for chemical bond breaking and therefore less non-linear phototoxicity. Furthermore, for thin cell cultures, we estimate a 2-fold lower temperature elevation when comparing the short time temperature rise (point scanning case) with the long time equilibrium temperature in wide-field 2P-excited fluorescence. Thus, our findings suggest that linear and non-linear phototoxicity is significantly reduced in 2P-RIM. Finally, our analysis also implies that wide-field 2P-excited fluorescence microscopy can be scaled using laser sources with increased average output power to investigate nearly arbitrary large areas without affecting the peak energy density nor increasing the equilibrium temperature at the sample.

Yet, this preliminary analysis, while encouraging, will need to be completed with further studies. The ratio of the temperature rise between the scanning and wide-field techniques, $\Delta T^A/\Delta T^a$, varies considerably with the sample thickness, heat conductivity, and scan speed. Furthermore, we have modeled the temperature rise in the wide-field configuration for uniform illumination. The analysis of the heating process in the speckle illumination scheme is more complex as the illumination is non homogeneous and the positions of the bright spots are rapidly changing. Nevertheless, we expect the linear heating in the RIM configuration to be less important than in the homogeneous illumination scheme because the speckle illumination generate more 2P-excited fluorescence photons than a homogeneous illumination for the same average power at the sample.

## 4. 2P-RIM App: description of the user interface

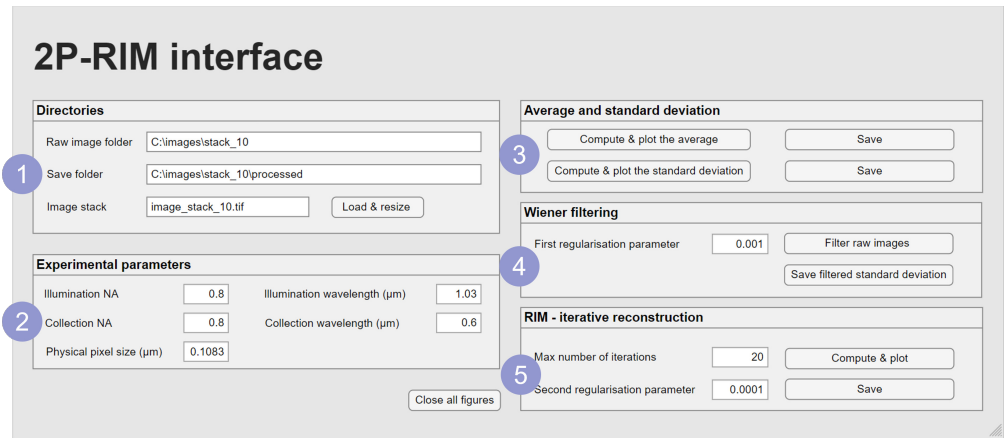

Fig. 1: 2P-RIM image processing interface. 1) Select the folders and images to process. 2) Choose parameter values corresponding to experimental conditions. 3) Basics: compute and save the average and standard deviation of the raw image stack. 4) Apply a Wiener filter to the raw images. 5) Perform the RIM iterative reconstruction on the images.

To ease the dissemination of 2P-RIM imaging we developed a standalone app for processing raw speckle illuminated 2P excited fluorescence images. The graphical user interface (GUI) is shown in SubFig. 1. Within the GUI, there are 5 sub-panels.
Panel-1 Directories: The direction and name of the raw images as well as the saving directory of the resulting images needs to be specified. Note that the input data must be tiff files (.tif). The processed images will be saved as tiff. Pressing the "Load & resize" button loads and resizes by a factor of 2 the stack of raw images.
Panel-2 Experimental parameters: The illumination and collection NA and wavelength needs to be entered. These values are required to compute the optical transfer functions, the wiener filter and the gradient for enhancing the image resolution within the iterative image reconstruction. Furthermore, the physical pixel size is specified i.e. the camera pixel dimensions in micrometers divided by the microscope's magnification factor.

Panel-3 Average and standard deviation image: The average and standard deviation image is computed & displayed and can be saved.

Panel-4 Wiener Filtering: Performs a Wiener deconvolution of the resized raw speckle images. The Wiener regularization parameter needs to be entered by the user and depends on the noise level of the experimental data. Typical values are $10^{-2} - 10^{-5}$. Pressing "Apply filter to raw images" shows the first raw and the standard deviation image after filtering.

Panel-5 iterative reconstruction: The image reconstruction requires information about the maximum number of iterations and a second regularization parameter. The wiener filter (first regularization) parameter to process the raw images will be read out from panel-4. Typical parameter values for all panels can be found within Fig. 1.

## 5. Determination of 2P-RIM's transverse resolution and optical sectioning

In this section, we study the resolution and optical sectioning of 2P-RIM with experiments on calibrated samples and with simulations.

### 5.1. Experimental study

We considered a sample made of small fluorescent beads of diameter 200 nm in water, emitting in the $\lambda \in [540 - 560]$nm range. We recorded 933 low resolution speckled images with a water objective of NA = 1.15 and excitation wavelength $\lambda_{\text{ex}} = 1030nm$. Fig. 2a shows the 2P-average and 2P-RIM images obtained from the speckled images. The full-width at half maximum (FWHM) of the intensity profile of a single bead is 402nm and 277nm for the 2P average and 2P-RIM techniques respectively. To estimate the FWHM of the point spread function, the size of the beads (200 nm) must be taken into account. Since both 2P-average and 2P-RIM images are linearly linked to the fluorescence density (a major asset of RIM reconstruction procedure is to keep the linearity with the sample brightness although the RIM data process involves non-linear transformations [3, 4]), the images can be modeled as the convolution of a point spread function (PSF) with the fluorescence density. For this purpose, we approximate the PSF as well as the fluorescence distribution of the beads by Gaussian functions. The FHWM of the fluorophore density is $\sqrt{3/2}D$ with $D$ being the diameter of the beads. The FHWM of the PSF can be estimated as $\text{FWHM}_{\text{PSF}} = (\text{FWHM}^2_{\text{image}} - \text{FWHM}^2_{\text{object}})^{1/2}$. Using this formula and $D$=173nm, results in a point spread function FWHM of 363nm and 216nm for the 2P-average and 2P-RIM images, respectively. The resolution gain brought by 2P-RIM is about 1.68. This achievement was confirmed with a sample made of actin fibers labeled with Alexa Fluor 568, Fig. 2b. Having identified a single, thin actin fiber, we measured a transverse FWHM of 529nm and 313nm within the 2P-average and 2P-RIM images which corresponds to a resolution gain of 1.69.

To determine the axial resolution of 2P-RIM, we examined a thin fluorescent layer, which was created by evenly painting one side of a coverslip with a permanent marker (orange Sharpie Neon). We moved the fluorescent plane away from the focal plane in steps of 0.5µm. At each $z$ position, we formed a 2P-average and a 2P-RIM image and integrated their pixel intensity over the whole field of view (about 200× 200 µm$^2$). We plot the 2P-average and 2P-RIM integrated intensity as a function of the distance to the focal plane in Fig. 2c. We observed no visible axial sectioning over a distance of 15µm for the 2P-average signal. In contrast, the 2P-RIM signal exhibited axial sectioning with a FWHM of about 2µm. The background signal of 0.5 that is visible in the 2P-RIM curve away from the focal plane is due to the camera dark noise variance.

### 5.2. Numerical study of 2P-RIM's performances

To confirm the resolution gain brought by 2P-RIM, we have carried out numerical simulations as close as possible to the experimental data. We considered a microscope with NA = 0.8 and fluorophores emitting at 515 nm yielding a point spread function $H$ of FWHM 390 nm. The excitation wavelength was set to 1030nm for all simulations.

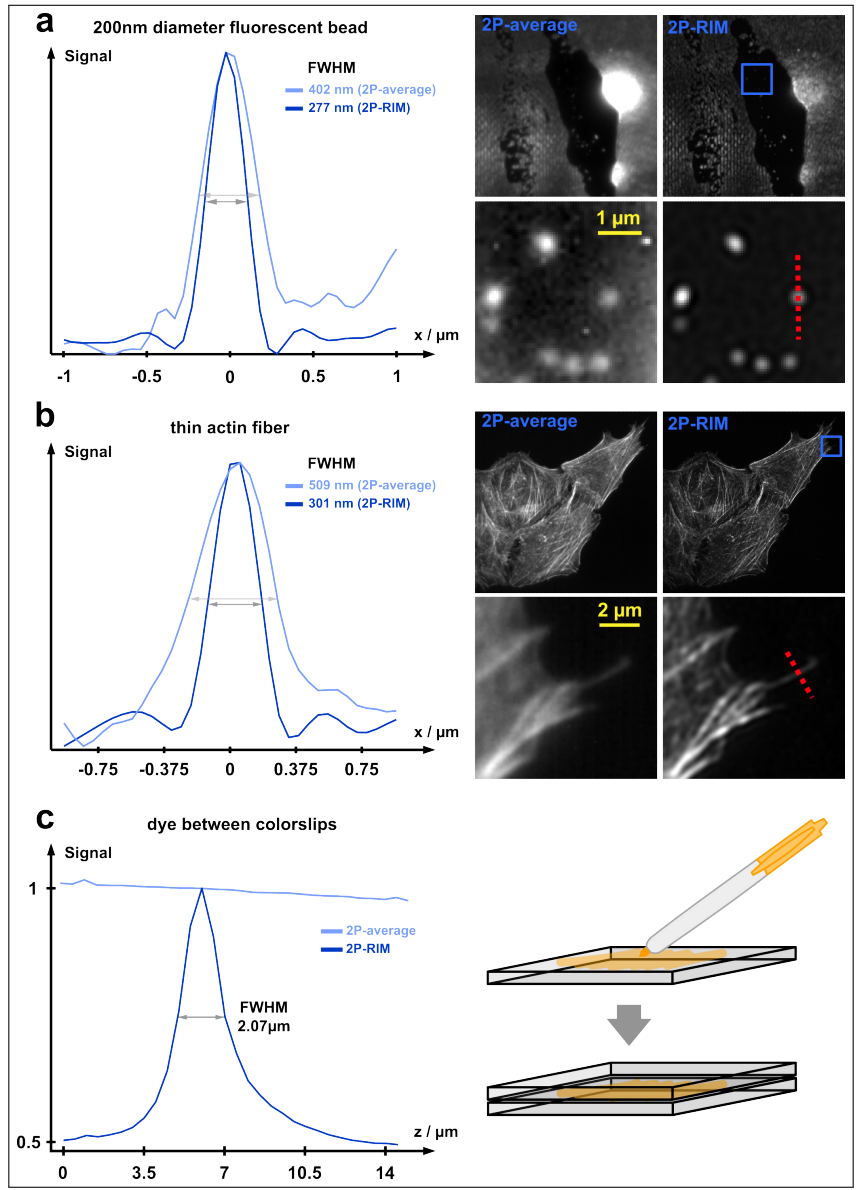

Fig. 2: Quantification of 2P-RIM's transverse and axial resolution on calibrated samples: **a)** 2P-average and 2P-RIM images of yellow fluorescence beads of 200nm diameter, large field of view (27µm) and zoom in over the blue square. The plot shows the intensity profiles along the red dotted line. The full-width at half maximum (FWHM) of the intensity profile of a single fluorescent bead was 402nm and 277nm for the 2P-average and 2P-RIM images, respectively corresponding to a resolution gain of 1.45. Accounting for the 200nm diameter of the beads yields effective PSFs with FHWM of 363nm and 216nm for the 2P-average and 2P-RIM image corresponding to a transverse resolution gain of 1.68. **b)** 2P-average and 2P-RIM images of thin actin fibers are displayed. The FWHM of the size of an actin fiber was found to be 509nm and 301nm within the 2P-average and 2P-RIM image, respectively, corresponding to a resolution gain of 1.69. **c)** The axial resolution of 2P-RIM. The sample is a quasi-homogeneous fluorescent plane, namely a stroke of a permanent yellow marker sandwiched between two coverslips. The sample was translated in 0.5µm steps along the axial direction. The plot shows the sum of the pixel intensities of the 2P-average and 2P-RIM images obtained for each $z$ position. 2P-RIM achieves an axial resolution of approximately 2µm while the 2P-average intensity profile shows no axial sectioning over a distance of 15µm.

### 5.2.1. Transverse resolution

To study the transverse resolution, we considered a resolution target consisting in a thin spoke pattern with fluorescence density $\rho(\varphi) = 1 + \text{sgn}\left[\cos\left(10.4\pi\varphi\right)\right]$ (where $\phi$ is the azimuth) discretized on a $601 \times 601$ square grid of step 13.5 nm. We simulated 2000 random speckled illuminations following Eq. (3) and formed 2000 speckled images following Eq. (1). The data were deteriorated with Poisson noise with the brightest pixel receiving 500 photons. The ground truth, 2P-average, standard deviation and 2P-RIM reconstruction of these synthetic data are shown in Fig. 3a. As expected, 2P-RIM improves significantly the transverse resolution (red, blue, and green dashed half circles show different diameters to guide the eyes). By comparing the radial Fourier support of the 2P-image and 2P-RIM images, we estimate the resolution gain brought by 2P-RIM to 1.6, comparable to that obtained experimentally. It should be noted that RIM resolution improvement depends on the frequency content of the observation point spread function $H$ and of the speckled excitation $S^2 = |E|^4$ [3]. In 2-photon microscopy, the Fourier support of $H$ roughly matches that of $S^2$ as in 1P RIM microscopy. However, the spectrum of the speckle square intensity, $S^2$, decays rapidly with the frequency (significantly more rapidly than the spectrum of speckled intensity $S$ which corresponds to the excitation in 1P RIM). As a result, one can consider that the 'effective' Fourier support of $S^2$ is smaller than that of $H$. This explains why we do not recover the two-fold improvement of the resolution as in 1P microscopy. By artificially diminishing the excitation wavelength to 515 nm, we could enhance the high spatial frequencies of $S^2$ and recover the two-fold resolution improvement, see Fig. 3a.

### 5.2.2. Optical sectioning

Next, we checked the optical sectioning of 2P-RIM. We first consider the properties of the 2P speckled excitation. In theory, the speckle model given by Eq. (3) yields a field that spreads towards infinity in all three directions with the same statistical properties everywhere. In theory, 2P speckled excitation does not produce a 'physical' optical sectioning. Yet, in practice, the speckled beam is focused through the objective onto a given Field of View (FOV). In Fig. (3b), we plot simulations of the speckled square intensity when the beam is focused onto different FOVs. The simulation was performed by imaging through a 4f system a speckled beam that was cropped, at a conjugate focal plane, by a circular field diaphragm [8]. We observe that if the FOV size is about $A = 200$ µm, the 2P speckled excitation is statistically homogeneous over the whole volume of investigation (about 200µm×200µm×80µm). On the other hand, if we reduce the FOV size to $A = 40$ µm, we observe that the 2P speckled excitation is confined axially within 40 µm about the focal plane.

The relationship between FOV and axial confinement can be roughly estimated by assuming that the averaged (over many speckle realizations) intensity of the cropped random illumination can be modeled as the light intensity stemming from a patch of incoherent sources filling the field diaphragm. In the sample space, each point source generates a beam that focuses at the focal plane with an angular distribution limited by the NA of the objective. The intensity of this beam in a transverse plane at a distance $z$ from the focal plane can be approximated by a Gaussian of waist $w(z) \approx \text{NA} \times z$ [1]. The averaged speckle intensity at plane $z$ is given by the convolution of this beam intensity with the incoherent source distribution. If the latter is also a Gaussian of waist $D$, its light intensity at plane $z$ is roughly a Gaussian with waist, $D(z) = D + w(z)$. To ensure energy conservation, the maximum of the Gaussian should satisfy $P(z) = P(0)\left[D/D(z)\right]^2$. This decay of the intensity as one moves away from the focal plane yields a decay in the 2P excitation following $1/(1 + 2NAz/D)^2$. The axial extension of the 2P excitation volume is thus roughly equal to the FOV width, as observed in the plot of Fig. 3b. This axial confinement is interesting as it can be used to reduce the phototoxicity and the background noise of densely marked samples if necessary. Yet, it is too large to provide useful optical sectioning. The micrometric optical sectioning of 2P-RIM is due to the standard deviation processing [9]. To confirm the experimental result of Fig. 2b, we simulated the speckled images of a homogeneous fluorescent plane placed at different distances from the focal plane (with the speckled illumination of FOV 200 µm). In Fig. 3b we plot the standard variation of the speckled images as a function of the distance to the focal plane. We recover an optical sectioning close to 2 µm in agreement with the experiment, Fig. 2b.

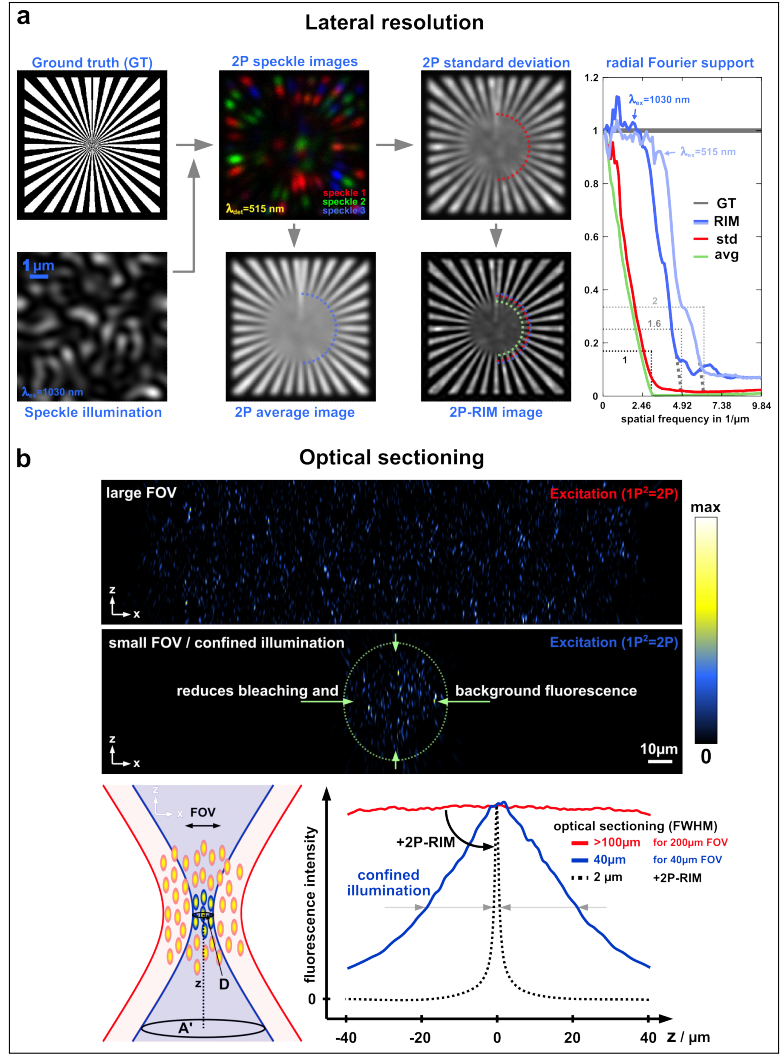

Fig. 3: 2P-RIM in silico: **a)** Lateral resolution improvement: a spoke pattern is illuminated by various speckle illuminations yielding a stack of 2P-excited fluorescence images (3 examples are shown as RGB). From this stack of 2P speckled images, the average image, standard deviation image and 2P-RIM reconstruction are computed. On the right, the object-normalized radial Fourier support of the average (avg), standard deviation (svg) and RIM image is displayed showing a resolution enhancement of a factor 1.6 at 1030nm excitation and 2 for an excitation at 515nm. **b)** Study of the optical sectioning. (x,z) cut in the sample volume (discretized over a cubic grid of 2400×2400×800 with voxel size 100nm) of the square intensity of a simulated speckled beam. The speckled beam was generated with N$A = 0.8$ and cropped by a field diaphragm of diameter 200 µm (top) and 40 µm (bottom). The axial confinement of the 2P excitation is clearly visible in the bottom case. The left-handed drawing illustrates the widening of the focused speckled beam which leads to the axial confinement of the 2P excitation. The right-handed plot provides a quantitative analysis of the optical sectioning. Images of a fluorescent plane under 2P speckled excitation are simulated for different positions $z$ of the plane. At each $z$, we calculate the pixel average and standard deviation of the speckled image (which is equivalent to the average over multiple speckled illuminations due to ergodicity). The red and blue curves correspond to the image average for speckled illuminations with 200 µm FOV and 40 µm FOV respectively. When the FOV is 40 µm, we observe a decay of the total emitted fluorescence as one moves away from the focal plane with FWHM about 40 µm. The black curve corresponds to the pixel standard deviation for the 200 µm FOV. The optical sectioning provided by the standard deviation process is about 2 µm.

## **References**

1. J. Mertz, *Introduction to optical microscopy* (Cambridge University Press, 2019).
2. J. W. Goodman, *Speckle phenomena in optics: theory and applications* (Roberts and Company Publishers, 2007).
3. S. Labouesse, J. Idier, M. Allain, G. Giroussens, T. Mangeat, and A. Sentenac, "Superresolution capacity of variance-based stochastic fluorescence microscopy: From random illumination microscopy to superresolved optical fluctuation imaging," Phys. Rev. A **109**, 033525 (2024).
4. T. Mangeat, S. Labouesse, M. Allain, A. Negash, E. Martin, A. Guénolé, R. Poincloux, C. Estibal, A. Bouissou, S. Cantaloube, E. Vega, T. Li, C. Rouvière, S. Allart, D. Keller, V. Debarnot, X. B. Wang, G. Michaux, M. Pinot, R. L. Borgne, S. Tournier, M. Suzanne, J. Idier, and A. Sentenac, "Super-resolved live-cell imaging using random illumination microscopy," Cell Reports Methods **1**, 100009 (2021).
5. G. Baffou and H. Rigneault, "Femtosecond-pulsed optical heating of gold nanoparticles," Phys. Rev. B **84**, 035415 (2011).
6. T. Lister, P. A. Wright, and P. H. Chappell, "Optical properties of human skin," J. Biomed. Opt. **17**, 0909011 (2012).
7. M. Tung, M. Trujillo, J. L. Molina, M. Rivera, and E. Berjano, "Modeling the heating of biological tissue based on the hyperbolic heat transfer equation," Math. Comput. Model. **50**, 665–672 (2009).
8. S. Heuke, K. Unger, S. Khadir, K. Belkebir, P. C. Chaumet, H. Rigneault, and A. Sentenac, "Coherent anti-stokes raman fourier ptychography," Opt. Express (2019).
9. C. Ventalon and J. Mertz, "Quasi-confocal fluorescence sectioning with dynamic speckle illumination," Opt. Lett. **30**, 3350 (2005).